\documentclass[a4paper,fleqn,usenatbib]{mnras}
\usepackage{graphicx}	
\usepackage{amsmath}	
\usepackage{amssymb}	
\usepackage[dvipsnames]{xcolor}
\usepackage{txfonts}
\usepackage{xcolor}
\usepackage{arydshln}

\usepackage{comment}
\def\lsun{\rm\, L_{\odot}}
\def\msun{{\rm M_{\odot}}}

\def\be{\begin{equation}}
\def\ee{\end{equation}}

\def\del#1{{}}
\newcommand\mearth{{\,{\rm M}_{\oplus}}}
\newcommand\mj{{\,{\rm M}_{\rm J}}}

\newcommand{\bdf}[1]{{ #1}}

\title[ALMA constraints on Core Accretion]{ALMA constraints on assembly of Core Accretion planets}

\author[Nayakshin et al.]{Sergei Nayakshin$^{1}$\thanks{sn85@le.ac.uk}, Vardan Elbakyan$^{1,2}$ and Giovanni Rosotti$^1$ \\
$^1$ School of Physics and Astronomy, University of Leicester, Leicester, LE1 7RH, UK.\\
$^2$ Institute of Astronomy, Russian Academy of Sciences, Pyatnitskaya str. 48, Moscow 119017, Russia}

\date{Accepted XXX. Received YYY; in original form ZZZ}

\pubyear{2022}

\begin{document}
\label{firstpage}
\pagerange{\pageref{firstpage}--\pageref{lastpage}}
\maketitle

\begin{abstract}
Resolved dust continuum and CO line ALMA imaging, and in some cases detection of H$\alpha$ emission, \bdf{hint} that young massive planets are abundant at wide separations in protoplanetary discs.  Here we show how these \bdf{observations} can probe  the runaway phase of planetary growth in the Core Accretion theory. Planets in this phase have the right range of masses to account for the predominantly moderate contrast gaps and rings seen in ALMA observations. However, we find that these planets gain mass and migrate inward very rapidly. As a result, the phase when they could produce gaps with properties similar to those observed is very short, i.e., $t_{\rm gap} \lesssim 0.1$~Myr, independently of the disc viscosity parameter. This would require many tens to hundreds of gas giant planets to be born per ALMA system, violating  the available mass budget of solids in realistic discs. This also predicts preponderance of discs with very wide gaps or complete inner disc holes, which is not observed. We show that suppression of both planet accretion and migration by a factor of at least ten is a possible solution to these serious problems. Future population synthesis models of planet formation \bdf{should aim to} address both exoplanetary data of older discless \bdf{planetary} systems and ALMA \bdf{discs with embedded planets in one framework}.
\end{abstract}


\section{Introduction}

A planet formation theory is a set of equations  describing evolution of planet variables, such as its mass $M_{\rm p}$, separation $a$, metallicity $z_{\rm p}$, etc., inside the parent protoplanetary disc. Protoplanetary disc evolution is itself a field of intense research \citep{WilliamsCiezaR11,AlexanderREtal14a}, with data indicating a tremendous variety in protoplanetary disc properties even for stars of similar mass, age, and environment \citep[e.g.,][]{ManaraT-etal17,Andrews20-Review}. Furthermore, exoplanetary data are very strongly dominated by planets orbiting  stars much older than $\sim 10$~Myr \citep[e.g.,][]{WF14}. These stars have lost all signs of their primordial discs from which they grew. This makes it impossible to ascertain the properties of the disc in which a given planet formed.  We also face similar uncertainties for the planets in the Solar System. Therefore, in the last several decades, to compare planet formation model predictions to observations theorists had to make do with statistical assumptions about the disc structure and parameters, and then run their models until disc dispersal \citep[e.g.,][]{PollackEtal96,IdaLin04a,MordasiniEtal12}. This situation is less than ideal, leaving much wiggle room for planet formation theories to contend with the data.

The recent protoplanetary disc imaging revolution \citep[e.g.,][]{BroganEtal15} is a game changer.
Instead of making statistical guesses about the properties of protoplanetary discs we can now constrain these from dust disc morphology, grain size, molecular line disc imaging, gas accretion rate onto the star, etc. Unlike exoplanets orbiting discless stars, ALMA candidate planets \citep[e.g.,][]{BroganEtal15,Dsharp1,LongEtal18,Pinte20-Dsharp-Vkinks} must be evolving right now, rather than in their long forgotten past, promising more direct observational constraints on the functions $F_i$.

In this paper we illustrate the power of this approach by focusing on a key step in the  Core Accretion (CA) scenario for planet formation \citep[e.g.,][]{Safronov69,PollackEtal96}. In {\em all} current flavours of CA, when a planetary (solid) core reaches the critical mass of $\sim 10\mearth$ while still embedded in the gaseous protoplanetary disc, a very rapid -- and thus called runaway -- gas accretion onto the planet sets in \citep[e.g.,][]{Mizuno80,Stevenson82,IkomaEtal00}. The runaway terminates when the planet reaches $\sim$ Jovian mass and opens a deep gap in the surrounding gaseous disc \citep{BateEtal03}. For a population of growing planets, very few planets are in the runaway phase at any given time; most are either too low mass or already too massive to be accreting gas rapidly. Accordingly, CA theories predict a pronounced planet  {\em desert} in the runaway mass range, that is, from $\sim$ twice Neptune ($0.1 \mj$) to $\sim$ Jupiter ($1 \mj$) masses\footnote{Note that the gap opening mass scales roughly as $\propto (H/R)^3$. Since discs are geometrically thicker at larger radii, the maximum planet mass in the CA scenario increases with separation from the star.} \citep{IdaLin04a,IdaLin04b}, although the amplitude of the desert depends on model detail \citep[e.g.,][]{MordasiniEtal09a,MordasiniEtal09b}.

Early analysis of radial velocity CORALIE/HARPS exoplanetary data appeared to confirm this theoretical prediction \citep{MayorEtal11}. More recently, microlensing observations \citep[e.g.,][]{SuzukiEtal16,SuzukiEtal18,Jung21-microlensing-PMF} and re-analysis of the CORALIE/HARPS sample \citep{Bennett21-No-runaway-desert} indicated instead a single power-law mass function from $\sim 10\mearth$ into the Jovian regime. Furthermore, modern 3D numerical simulations find much smaller gas accretion rates onto planets \citep[][]{Szulagyi14,OrmelEtal15,FungChiang16,LambrechtsLega17,LambrechtsEtal19-No_CA-runaway} than previous 1D studies found.

Using the census of planets growing in their discs compiled by \cite{LodatoEtal19}, \cite{NayakshinEtal19} showed that ALMA planets must be growing $\sim$ an order of magnitude slower than CA scenario predicts. \cite{NayakshinEtal19} used idealised power-law gas-only models for ALMA discs and neglected planet migration.  In this paper we extend this work in several significant ways. We include planet migration, and calculate the time-dependent structure of discs via a 1D model that includes a self-consistent gravitational torque exchange between the disc and the planet, including disc gap opening. We also include gas accretion onto planets by explicitly removing mass from the disc, thus combining constraints from planet accretion and migration. Finally, dust dynamics in the disc is solved for, enabling analysis of time evolution of planet-induced dust gaps, inner holes, and rings.

Our paper is organised as following. In \S 2 we introduce our methods, in \S \ref{sec:HD_SS} we apply these methods to a well known protoplanetary disc in HD163296 to illustrate how planets can be used to constrain CA scenario. In \S 4 we apply these ideas to a sample of ALMA gaps, and in \S 5 we present a discussion of main results of our paper.

\section{Method}\label{sec:methods}

\subsection{Gas, dust and planet dynamics}\label{sec:dynamics}

Due to numerical limitations, population synthesis modelling of planet formation has so far only been performed in 1D \citep[e.g.,][]{IdaLin04a,MordasiniEtal09a,ColemanNelson14,NayakshinFletcher15,NduguEtal19}. For similar reasons, and to be directly comparable with previous work, we follow this approach here, employing 1D geometrically thin azimuthally averaged disc model.

We are interested in the outer regions of protoplanetary discs with accretion rates usually well below $10^{-7}\msun$~yr$^{-1}$. Under such conditions the discs are passively heated by irradiation from the central star, and the midplane temperature scale with radius $R$ approximately as $T\propto R^{-1/2}$ \citep{DAlessioEtal01}. We therefore set
\begin{equation}
     T = T_0 \left(\frac{R_0}{R}\right)^{1/2}
    \label{T-profile}
\end{equation}
where $R_0 = 100$~AU and $T_0$ is specific for each source we model. The disc geometric aspect ratio is $H/R = h_0 (R/R_0)^{1/4}$, where $h_0 = (k_b T_0 R_0/GM_*\mu)^{1/2}$, $k_b$ is the Boltzmann's constant and $\mu = 2.4m_p$ is the mean molecular weight, and $M_*$ is the stellar mass.

Time evolution of the gaseous disc surface density is described by the standard viscous model, which reads
\begin{equation}
  \frac{\partial\Sigma}{\partial
  t} =\frac{3}{R}\frac{\partial}{\partial R}
  \left[R^{1/2}\frac{\partial}{\partial R}(R^{1/2}\nu\Sigma)\right]
  -\frac{1}{R}\frac{\partial}{\partial
  R}\left(R v_\lambda\Sigma\right)  - \dot\Sigma_{-}\;.
\label{eq:dSdt0}
\end{equation}
In this equation, the terms on the right are: the viscous diffusive mass transfer, the radial mass flow due to the tidal torque from the planet, and the mass loss term due to accretion onto embedded objects. The velocity $v_{\lambda} = 2\lambda_t/(R\Omega_{\rm K})$ may be thought as the velocity imposed on the gas by the specific tidal torque $\lambda_t$ from the planet. The integral of the tidal torque over the whole disc gives the total rate of the angular momentum gain of the disc due to planet-disc interaction, $\Lambda_{\rm t}$ (cf. eq. \ref{lambda_t_int}). We note in passing that $\Lambda_{\rm t}$ can be positive (the planet migrates inward) or negative (the planet migrates outward). The conservation of the angular momentum in the disc-planet system demands that the planet-star separation $a$ evolves as
\begin{equation}
    \frac{d a}{dt} = - \frac{2 \Lambda_{\rm t}}{M_{\rm p}} \left(\frac{a}{GM_*}\right)^{1/2} = - \frac{a}{t_{\rm mig}}\;,
    \label{dadt0}
\end{equation}
where we defined the planet migration time, $t_{\rm mig}$. We use the \cite{CridaEtal06} parameter 
\begin{equation}
C_{\rm p} = {3H\over 4R_{\rm H}} + 50 \alpha_{\rm v} \left({H \over a}\right)^2 {M_*
  \over M_p} 
\label{CridaP0}
\end{equation}
to differentiate between type I planet migration (no deep gap in the gas surface density at the location of the planet), $C_{\rm p} \geq 1$, and type II regime (a deep gap is opened), $C_{\rm p} \leq 1$. In the type I regime we use \cite{Paardekooper11-typeI} expressions for $\Lambda_{\rm t}$, whereas in type II regime the tidal torque is an integral (cf. eq. \ref{lambda_t_int}) over the specific torque $\lambda_t$, given by the widely used expression in the literature, \citep{LinPap86,armibonnell02,LodatoClarke04,AlexanderEtal06,LodatoEtal09,DipierroLaibe17}:
\begin{eqnarray}
\label{eq:torque}
\lambda_{t} = & \displaystyle f  \frac{q^2}{2} \left(\displaystyle
\frac{a}{\Delta R}\right)^4 \frac{G M_*}{a} & \text{ for }
R>a \\
\nonumber \lambda_{t} = & - f  \displaystyle \frac{q^2}{2}\left(\displaystyle 
\frac{R}{\Delta R}\right)^4 \frac{G M_*}{a} & \text{ for } R<a\;,
\end{eqnarray}
where $\Delta R =
|R-a|$ and the factor $f$ is a constant set to $f=0.5$ here. The torque term diverges at the planet location, and hence one  smooths it for
$|\Delta R| <\max[H,R_H]$ \citep[e.g., ][]{SyerClarke95}. As a number of authors in the past we use a prescription to transition smoothly from type I to type II planet migration (see appendix  \ref{sec:app_numerics} for detail). 

The protoplanetary disc viscosity is given by the \cite{Shakura73} model, $\nu = \alpha c_s H$, where $c_s = (P/\rho)^{1/2}$ is the isothermal sound speed, with $P$ and $\rho$ being the disc midplane pressure and density.

We neglect dust back reaction on the gas and assume that dust radial velocity $v_{\rm d}$ is given by the terminal velocity approximation plus the diffusion term \citep[see, e.g.,][]{RosottiEtal16}. The dust surface density then obeys the mass continuity equation, modulo mass loss due to planet accretion (when included):
\begin{equation}
  \frac{\partial\Sigma_{\rm d}}{\partial
  t} =
  -\frac{1}{R}\frac{\partial}{\partial
  R}\left(R v_{\rm d}\Sigma_{\rm d}\right) - \dot \Sigma_{\rm d}\;,
\label{eq:dust_dt0}
\end{equation}
where 
\begin{equation}
    \dot \Sigma_{\rm d} = \frac{\Sigma_{\rm d}}{\Sigma} \dot\Sigma_{-}\;.
    \label{mdot_dust0}
\end{equation}
The latter equation merely states that dust is carried with the gas into the planet during the runaway gas accretion phase onto it. This is appropriate since our dust particles are usually in the small Stokes number regime. We do not include pebble accretion onto our planets as this is important only for planets before they enter the gas runaway phase \citep[e.g.,][]{NduguEtal19} and we are not concerned here with how exactly our planets gained enough solids (via planetesimals or pebbles) to get into the runaway regime. We solve eqs. \ref{eq:dSdt0} and \ref{eq:dust_dt0} in an explicit manner.

Protoplanetary disc surface density is often modelled via a power-law with an exponential rollover, 
\begin{equation}
    \Sigma = \Sigma_0 \frac{R_0}{R} \exp\left[ 
    -\frac{R}{R_c}\right]\;,
    \label{sigmaIn-0}
\end{equation}{}
where $\Sigma_0$ is a normalisation constant, $R_{\rm c}$ is the critical radius  \citep[e.g., in HD163296 it is 150 AU, see][]{DullemondEtal20-HD163296}. The values of $R_{\rm c}$ for most of the observed discs are not well constrained, but fortunately are not particularly important for our modelling here. Dust continuum observations only constrain the outer edge of large dust particles disc \citep[e.g.,][]{Rosotti19-Opacity-Cliff}. The location of the outer edge of the gas disc is harder to pinpoint but is likely to be $\sim$ a few times that of the dust disc. Since the planets we are interested here are by definition within the dust disc, $R_{\rm c}$ is probably much larger than $a$ in these systems. 

For numerical convenience, and in line with a common approach in the field \citep[e.g.,][]{RosottiEtal16,Dsharp7} we use the following initial gas surface density profile:
\begin{equation}
    \Sigma = \Sigma_0 \frac{R_0}{R}\;,
    \label{sigmaIn-1}
\end{equation}
sharply cut at $R=R_{\rm out} = 200$~AU, where we place a reflecting boundary condition. These simplifications do not affect our main conclusions but are numerically convenient.
Coupled with $T(R)$ given by eq. \ref{T-profile}, this density profile results in a radially constant gas accretion rate in the disc, $\dot M$. To keep our disc in this state we further  keep the gas and dust surface densities at the outer boundary fixed, while allowing the dust and gas to freely pass though the inner boundary \citep[such boundary conditions are sometimes called ``evanescent"; ][]{Dsharp7}. 

\subsection{Planet accretion}\label{sec:accretion}

To model gas accretion term $\dot \Sigma_{-}$ onto our planets we follow the well known Bern model for planet formation \citep[e.g.,][]{Bern20-1,Bern20-2}. In the Bern model, the gas capture radius of the planet is defined as
\begin{equation}
    R_{\rm gc} = \frac{k_{\rm acc} R_{\rm H} R_{\rm acc}}{R_{\rm acc} + k_{\rm acc} R_{\rm H}}
    \label{Rgc0}
\end{equation}
where $R_{\rm acc} = G M_{\rm p}/c_s^2$ and $k_{\rm acc} = 1/4$. Gas accretion proceeds in the 3D regime for low mass planets, when $R_{\rm gc} \leq H$, and in 2D regime for high mass planets when $R_{\rm gc} > H$. The maximum (runaway) 2D gas accretion rate onto the planet in \cite{Bern20-1} model is
\begin{equation}
    \dot M_{2D} = 2 R_{\rm gc} \Sigma(a) v_{\rm rel}\;,
    \label{Mdot2d-0}
\end{equation}
where $v_{\rm rel} =$~max$[\Omega_K R_{\rm gc}, c_s]$. 

In the Bern model, the gap opening in the disc by the planet is not computed from first principles since the planet tidal torque is not applied to the disc while the planet is in the type I migration regime. The motivation for this approach is the fact that the planet affects the {\em gas} disc weakly except very close to it, so $\Sigma(a)$ in eq. \ref{Mdot2d-0} is approximately the unperturbed gas surface density. The Bern model then assumes that a deep gap in the disc emerges once $C_{\rm p}$ drops below unity. We note that this is a common approximation in population synthesis for planet formation \citep[e.g.,][]{ColemanNelson14}.

Due to the aims of our paper, however, gap formation in the gas disc must be explicitly resolved since dust is usually affected by the planets stronger than gas \citep{DipierroEtal16a}, so that even planets with $C_{\rm p} > 1$ can open deep gaps in the discs \citep{DipierroLaibe17}. Therefore we use the actual $\Sigma(R)$ in eq. \ref{Mdot2d-0}
rather than the unperturbed value. This reduces the accretion rate onto the planet when it starts opening a gap, making the main conclusions of our paper more conservative.

Since $\Sigma(R)$ may be changing rapidly near the planet location in our more physically complete model, to find the gas accretion rate onto the planet, $\dot M_{\rm acc}$, instead of using the unperturbed gas density in eq. \ref{Mdot2d-0}, we  take the following integral:
\begin{equation}
    \dot M_{\rm acc} = f_{\rm geom} \int_{a-R_{\rm H}}^{a + R_{\rm H}} dR \Sigma(R) v_{\rm gc}\;,
    \label{Mdot2d-1}
\end{equation}
where the gas capture velocity $v_{\rm gc} =$~max~$[c_s, \Omega_K |R-a|]$, and $f_{\rm geom} =$~min~$[R_{\rm gc}/H, 1]$ is the geometrical factor that is smaller than unity in the 3D accretion case.
For the case of no gap opened in the disc, within a factor of order unity, the result of eq. \ref{Mdot2d-1} is the same as eq. (47) in \cite{Bern20-1}. When the planet mass increases and the gap begins to open, however, eq. \ref{Mdot2d-1} self-consistently reduces gas accretion onto the planet, eventually terminating its growth and preventing the planet from accreting the whole gas disc. We further comment that in the Bern model the planets may be on eccentric orbits due to their interactions with other planets, which increases the feeding zone of the planet and may lead to it reaching  higher terminal masses \citep[cf. eqs. 49 \& 50 in][]{Bern20-1}. To be conservative in our conclusions, here we do not allow eccentric orbits. Relaxing this assumption would make the need to suppress gas accretion onto planets even stronger.

\section{An Example: the planet at 86 AU in HD163296}\label{sec:HD_SS}

To exemplify the range of results amid parameter dependencies we focus on one particular case before turning our attention to the \cite{LodatoEtal19} sample of candidate planets in ALMA discs. HD 136296 is one of the closest and therefore one of the best studied Herbig Ae stars, with mass $M_* = 2.04\msun$, luminosity $L_* = 17 \lsun$, and an estimated age of $t_*\approx 5$~Myr \citep{Dsharp1}. Its protoplanetary disc is one of a very few for which CO and DCO$^{+}$ molecular line observations with ALMA are of sufficiently high quality to set an upper limit to the disc turbulence parameter $\alpha_{\rm v}\lesssim 3\times 10^{-3}$ \citep{Flaherty17-HD163296-turb}. We set $T_0 = 65$~K at $R_0 =100$~AU \citep[Fig. 6 in][]{RabEtal20-Temperature-in-gaps}. Note that this temperature is found at the CO molecule emitting surface, and is larger than $T\approx 25$~K temperature derived for the midplane of the disc in this source by \cite{DullemondEtal20-HD163296}. However, by picking the larger of the derived temperatures here we obtain the most conservative results in the context of our vertically and azimuthally averaged modelling. As we show in Appendix \ref{sec:app_numerics}, the lower the disc temperature, the stronger the ALMA constraints on the Core Accretion scenario. This is because at lower disc temperatures gas accretion rate onto the planet is higher, the type I migration rate is faster at the same disc mass, and a deep gap opening in the gas disc occurs sooner.

The putative planet we shall focus on produces a well defined gap at $a_{\rm p} = 86$ AU, characterised in \cite{Dsharp7} by $\Delta_{\rm gap} = \Delta R/a = 0.17$, where $\Delta R$ is the gap radial width, and the gap/ring intensity contrast $C_{\rm gap} = 0.15$ \citep[for definitions see][ and \S \ref{sec:runaway_start} below]{DSHARP-6,Dsharp7} The case for this candidate planet has also been strengthened by the detection of velocity perturbations both in the vertical direction in the disc \citep{Teague19-HD163296} and in the disc plane
direction (velocity kink) visible through the CO molecular line observations \citep{Pinte18-HD163296,Pinte20-Dsharp-Vkinks}. The disc mass in this source has been estimated through dust mass conversion to gas (assuming 100 for the gas-to-dust ratio) and also through disc chemodynamical modelling at $M_{\rm disc} = (0.05-0.3) \msun$ \citep{QiEtal11,MuroA-Etal18-HD163296,BoothEtal19-HD63296}. Further, \cite{Powell19-DustLanes} used the dust radial drift arguments to estimate the gas mass in HD163296 at $M_{\rm disc} = 0.2\msun$. We therefore chose $\Sigma_0$ such that the gas mass of our disc is $0.2\msun$. We fix the inner and outer radii of our computational domain at 5 and 200 AU, respectively.

\subsection{Release from an artificial steady state}\label{sec:ss_start}

\begin{figure*}
\includegraphics[width=0.32\textwidth]{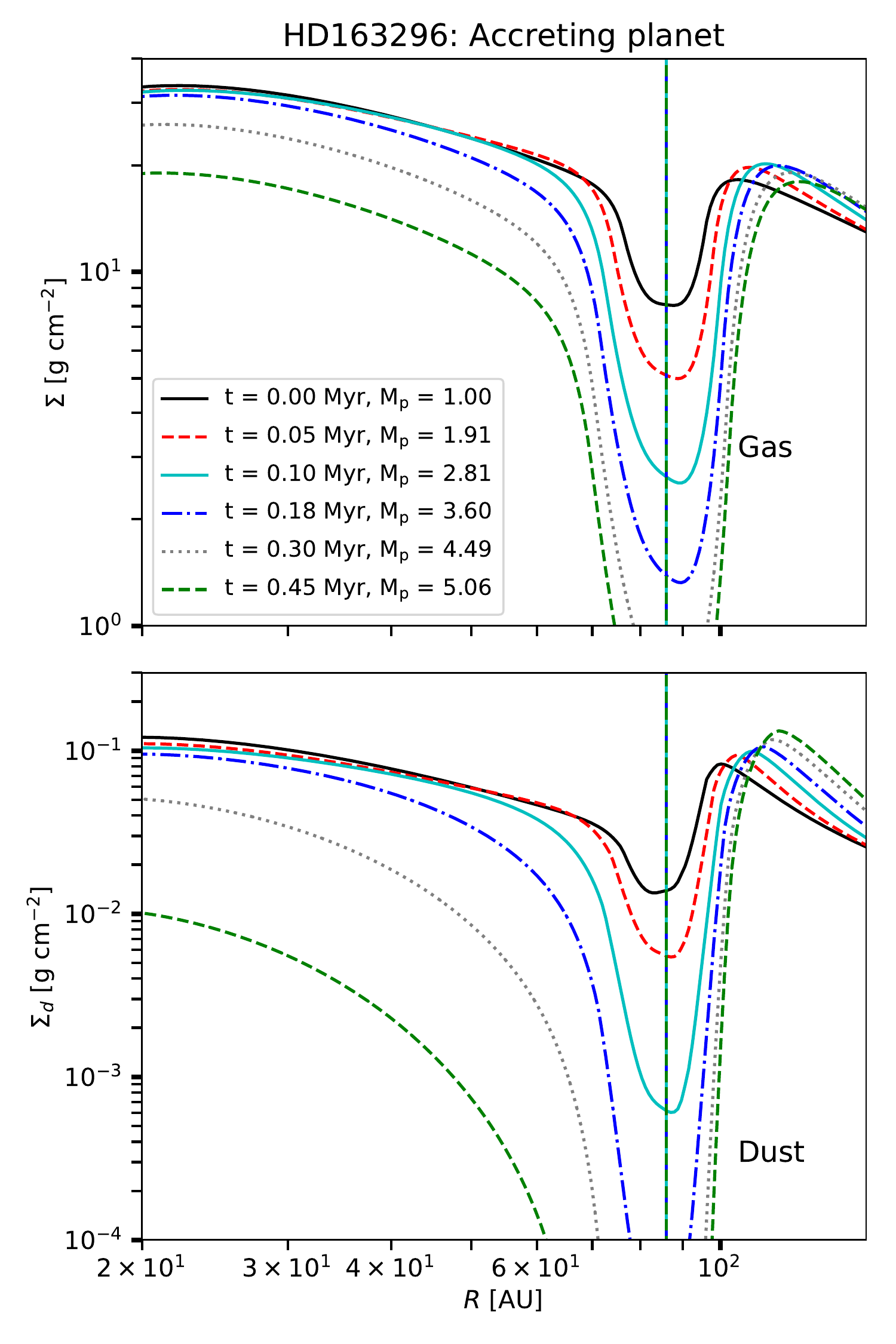}
\includegraphics[width=0.32\textwidth]{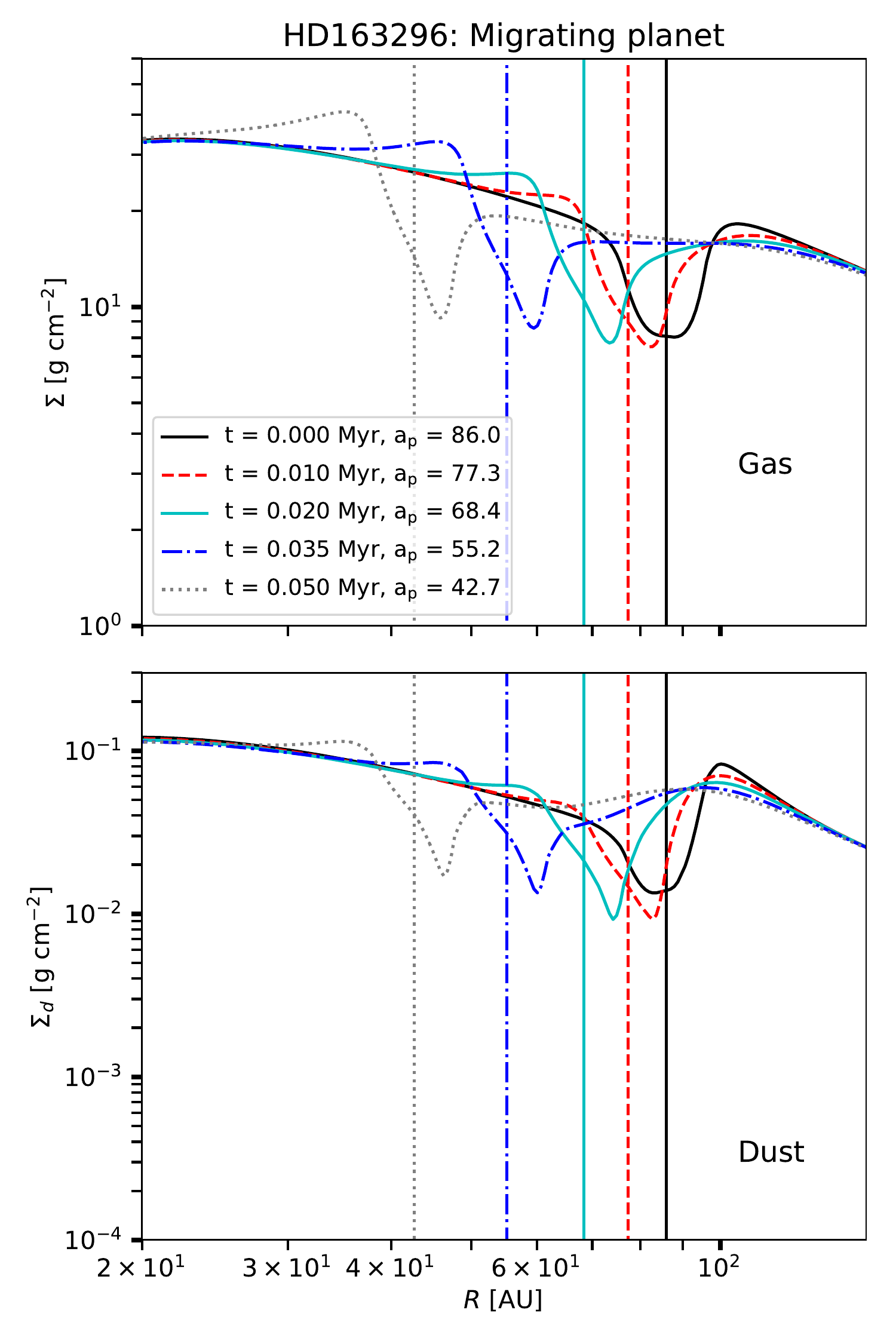}
\includegraphics[width=0.32\textwidth]{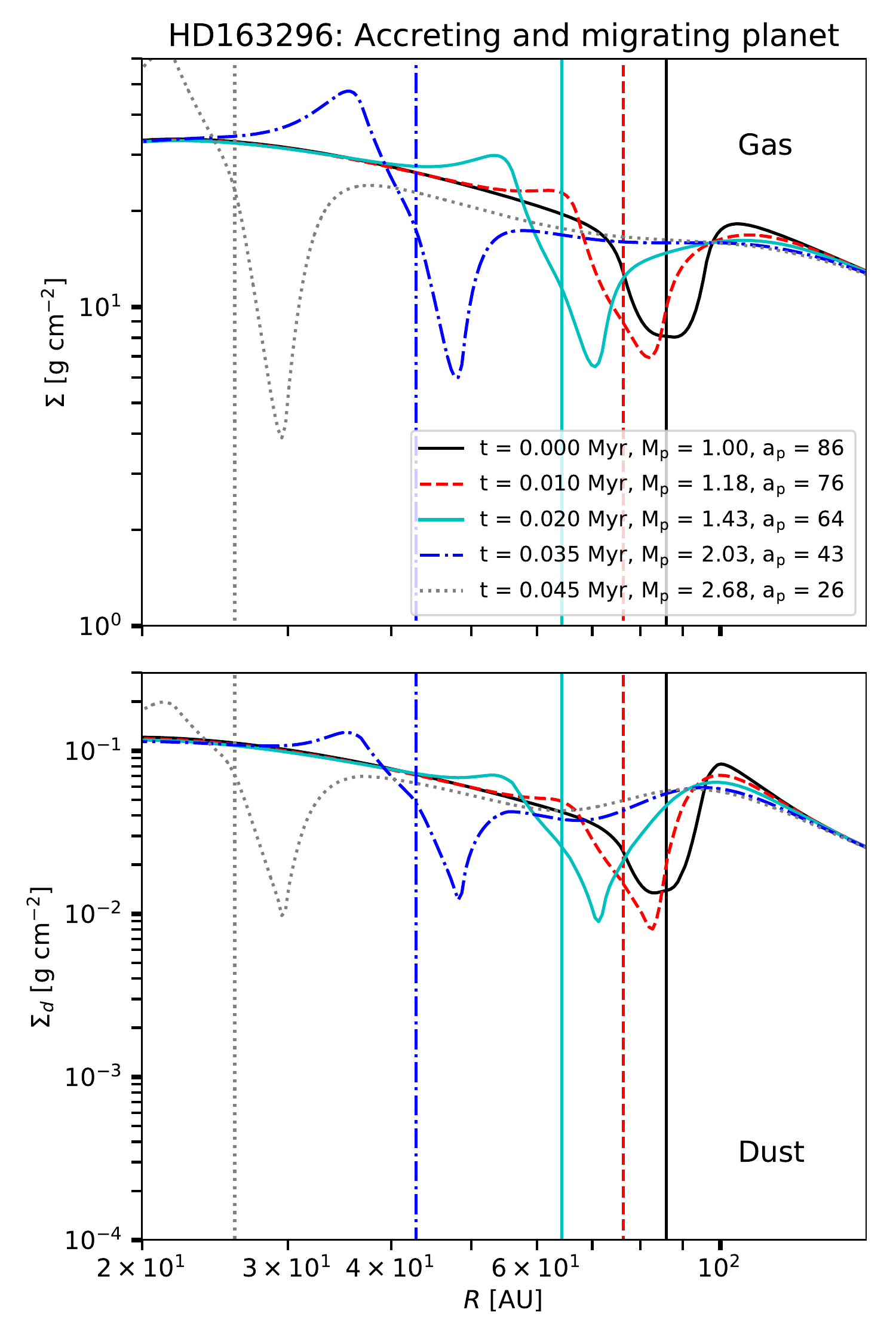}
\caption{Evolution of the planet-disc system when the planet is allowed to evolve according to one of the following: accrete; migrate; both accrete and migrate (from left to right). The vertical lines show the radial location of the planet at the times indicated in the legends.  The initial planet location and mass were found by searching for a good fit to the values of the gap width and the ring/gap contrast observed for the D86 gap in the disc of HD163296. The experiments here indicate that when the planet is allowed to evolve as expected it leaves this desirable mass-location parameter space worryingly quickly, in $\sim 0.03$ Myr. }
\label{fig:HD163296}
\end{figure*}

Since multi-dimensional numerical simulations of dusty discs with embedded planets are expensive, it is often assumed that planet mass and orbit remain constant during the simulations \citep[e.g.,][]{Dsharp7}. One then runs the gas-dust simulations for a range of planet masses, $M_{\rm p}$, disc viscosity parameter $\alpha_{\rm v}$, maximum grain size $a_{\rm max}$, etc.,  to try and zero in on model parameters that best reproduce the morphology of the ALMA dust continuum emission. In this section we follow a similar approach, placing the planet at $86$~AU. 

We start by setting the disc viscosity parameter $\alpha_{\rm v} =10^{-3}$; this parameter will be varied in the sections to follow. To reach a quasi steady state the simulations are run for 0.5 Myr. By performing a number of numerical experiments we found that grains with the maximum size $a_{\rm max} = 1.5$~mm and planet mass $M_{\rm p} \approx 1 \mj$ reproduce the observed values for the gap width and contrast in HD163296 well. This planet mass is reasonably close to those favoured by previous authors \citep[e.g.,][]{LiuEtal19}.
Once we determined the appropriate $M_{\rm p}$, we restart the simulation from the quasi steady state that the system reached at $t=0.5$~Myr, now allowing the planet to do one of the following: (i) accrete gas and dust as per \S \ref{sec:methods} but not migrate; (ii) migrate but not accrete gas; (iii) both accrete and migrate. 

Fig. \ref{fig:HD163296} shows the resulting gas (top panels) and dust (bottom panels) surface density profiles at several different times. The time in the legend is counted from restart, so $t=0$ corresponds to the quasi steady state. When only accretion is enabled (left panels) the planet evolves towards a gas gap opening planet at its (forced) location. This evolution is rapid, both in terms of planet mass and the dust surface density profile. The planet nearly doubles its mass in 0.05 Myr. The lower left panel shows that dust gap becomes an order of magnitude deeper within just 0.1 Myr. Within 0.5 Myr the planet grows in mass to about $5\mj$ and opens a complete dust gap from the star to $\sim 100$~AU. Even the gas accretion onto the star is affected by that point. Clearly such evolution turns the modest gap in this system into a transition like disc with a very large inner hole, somewhat like PDS70, except the planet is much further out in HD163296.

When only planet migration is allowed (middle panels) we observe that the dust gap parameters change less significantly but the position of the gap moves towards the star extremely rapidly. The planet-star separation shrinks  by a factor of two in about 0.03 Myr. This is slightly longer than analytical estimates for a planet migrating in the type I planet regime; the migration time in our disc is given by
\begin{equation}
    t_{\rm mig1} = \frac{1}{2\gamma \Omega} \frac{M_*^2}{M_{\rm p} \Sigma R^2} \left(\frac{H}{R}\right)^2 
    \label{t1-0}
\end{equation}
where $\Omega_{\rm K} = (GM_*/R^3)^{1/2}$ is the Keplerian angular frequency at the planet location. The dimensionless factor $\gamma=2.5$ \citep{PaardekooperEtal10a} is set by the assumed disc properties. Numerically,
\begin{equation}
    t_{\rm mig1} = 0.016 \hbox{ Myr}\; \frac{R}{100 \text{ AU}} \left(\frac{M_*}{ \msun}\right)^{1/2} \left(\frac{1 \mj}{M_{\rm p}}\right) \frac{0.1 \msun}{M_{\rm disc}} \;.
    \label{t1-1}
\end{equation}{}
For HD163296 and $R=86$~AU this yields $t_{\rm mig} \approx 0.01$~Myr. The fact that our planet migrates a factor of a few slower is probably due to its opening a moderately deep gap in the disc. 

The middle panel in Fig. \ref{fig:HD-one-run} also shows a change in the dust morphology from the ``ring behind the planet" to ``the ring inside the planet" one.  These changes in the dust surface density profile are consistent with the results of \cite{Meru19-Ring}. We shall discuss this further in \S \ref{sec:Fa10}.

Finally, when both accretion and migration are permitted the system evolution is even more rapid. As the planet gains mass its inward migration accelerates since it is migrating in the type I regime in which the migration time is proportional to the planet mass. For the parameters of this system the planet stops migrating rapidly only when it reaches $\sim 10$~AU as it enters the type II regime.

These experiments demonstrate that allowing the planet to evolve as predicted by the runaway CA scenario leads to rapid and very significant changes in the planet parameters and the dusty disc morphology. It appears that an evolutionary time scale for this disc-planet system is of order $0.05$~Myr, which is surprisingly short for a $\sim 5$ Myr old system.

\subsection{Release at the onset of runaway}\label{sec:runaway_start}

\begin{figure*}
\includegraphics[width=0.98\textwidth]{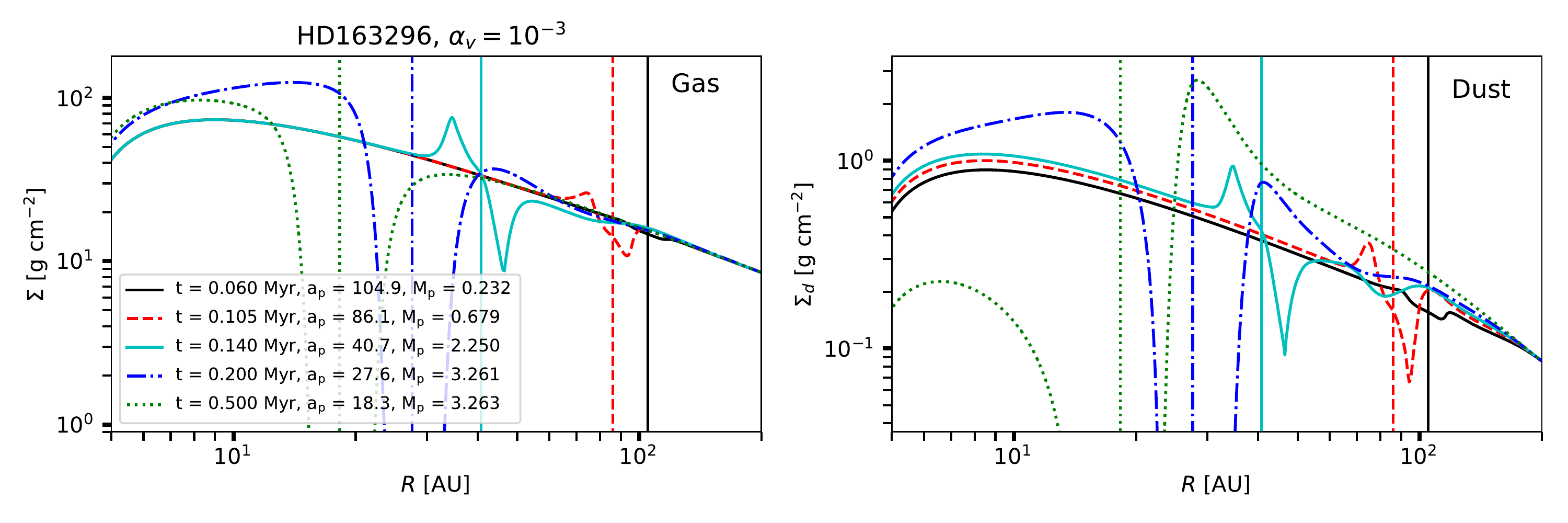}
\includegraphics[width=0.98\textwidth]{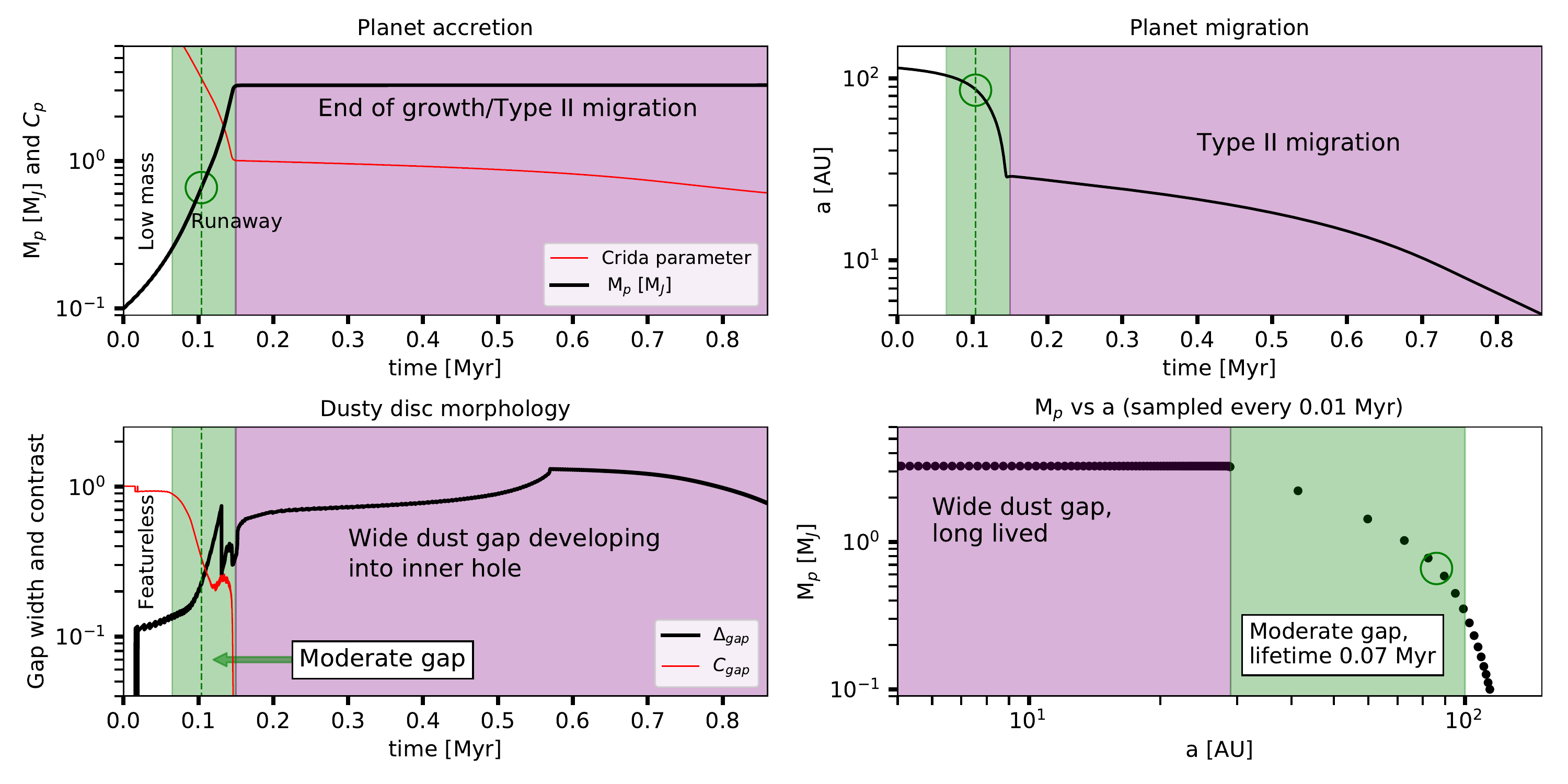}
\caption{Snapshots of the disc gas and dust profiles (top panels,  similar to Fig. \ref{fig:HD163296}), and the history of the  planet-disc system ( middle and bottom row of panels).  The middle row left and right panels show planet mass $M_{\rm p}$ and separation $a$ versus time, respectively, whereas the bottom row right panel presents $M_{\rm p}$ vs $a$. The bottom row left panel shows the evolution of two key gap parameters versus time, the gap width $\Delta_{\rm gap}$ and the gap contrast $C_{\rm gap}$ (eqs. \ref{Delta_gap0} and \ref{Cgap}, respectively). The different background colours in the middle and the bottom rows are to help the reader to locate the three characteristic regimes in the disc evolution, which often but not always have a relation to the state of the planet. In particular, in the earliest pre-runaway phase the planet is too low mass to open a significant gap, so the disc looks featureless (white background); in the runaway phase the planet gains mass quickly and the gap in the disc is moderate (green background); in the detached/end of growth phase the planet reaches its final mass, opens a deep gap in both gas and dust and migrates inward slowly in the Type II regime (purple background).} 
\label{fig:HD-one-run}
\end{figure*}

\begin{figure*}
\includegraphics[width=0.98\textwidth]{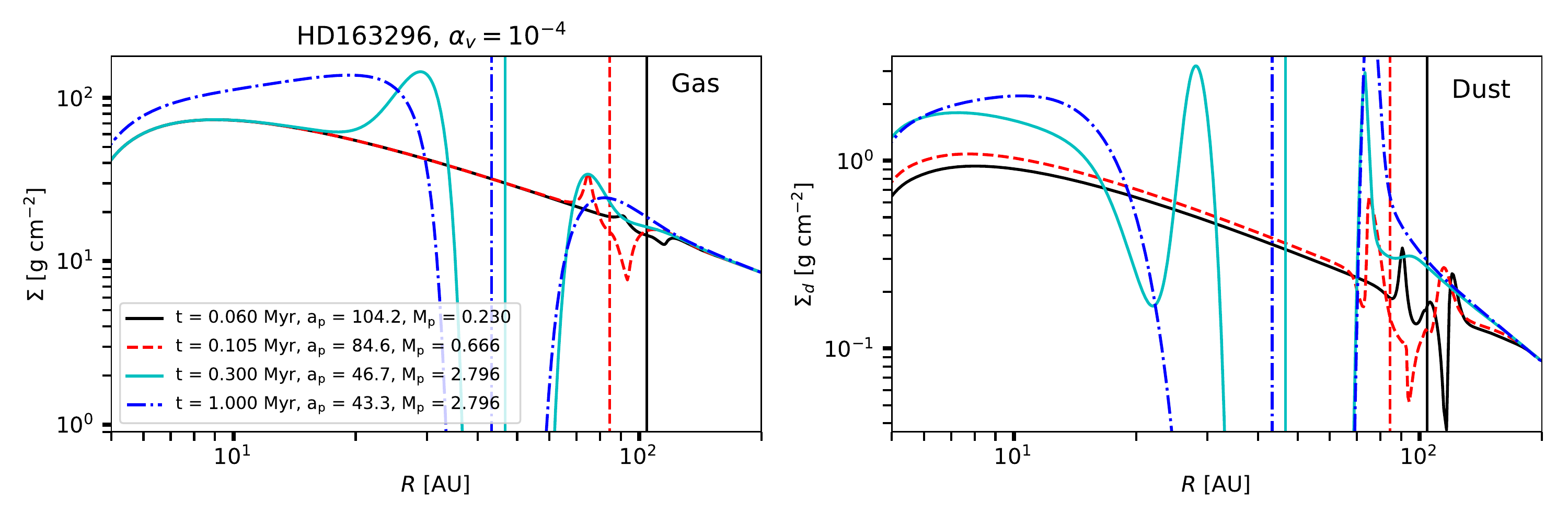}
\includegraphics[width=0.98\textwidth]{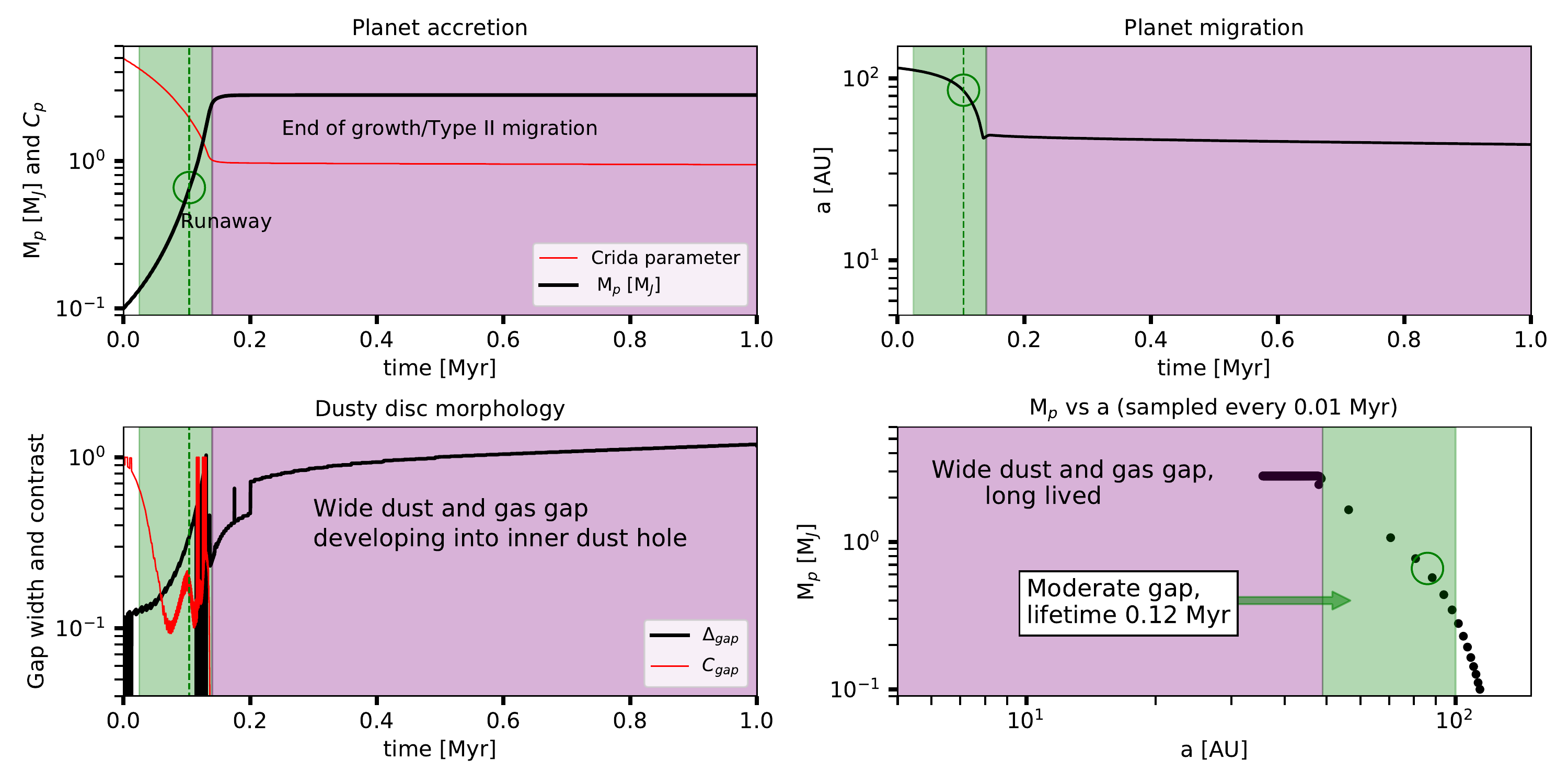}
\caption{Same as Fig. \ref{fig:HD-one-run} but for a lower viscosity parameter, $\alpha_{\rm v} = 10^{-4}$. Note that the planet now reaches the type II migration regime earlier and then migrates slower, so it spends a longer time at large separations. However, the dusty disc morphology in this case evolves into the observationally undesirable configuration with the gap too deep and too wide, even sooner. \bdf{Despite an order of magnitude lower value of $\alpha_{\rm v}$ than in Fig. \ref{fig:HD-one-run}, the duration of the moderate gap phase is similarly short,} $\sim 0.12$~Myr.}
\label{fig:HD-alpha1e4}
\end{figure*}

\begin{figure*}
\includegraphics[width=0.98\textwidth]{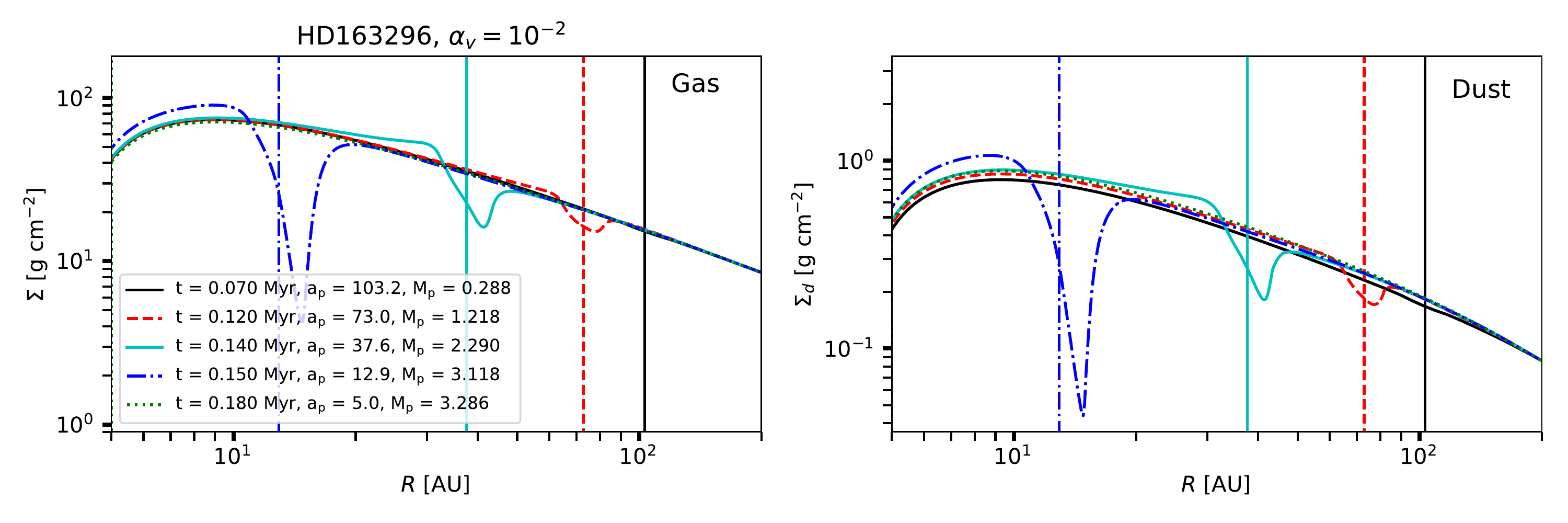}
\includegraphics[width=0.98\textwidth]{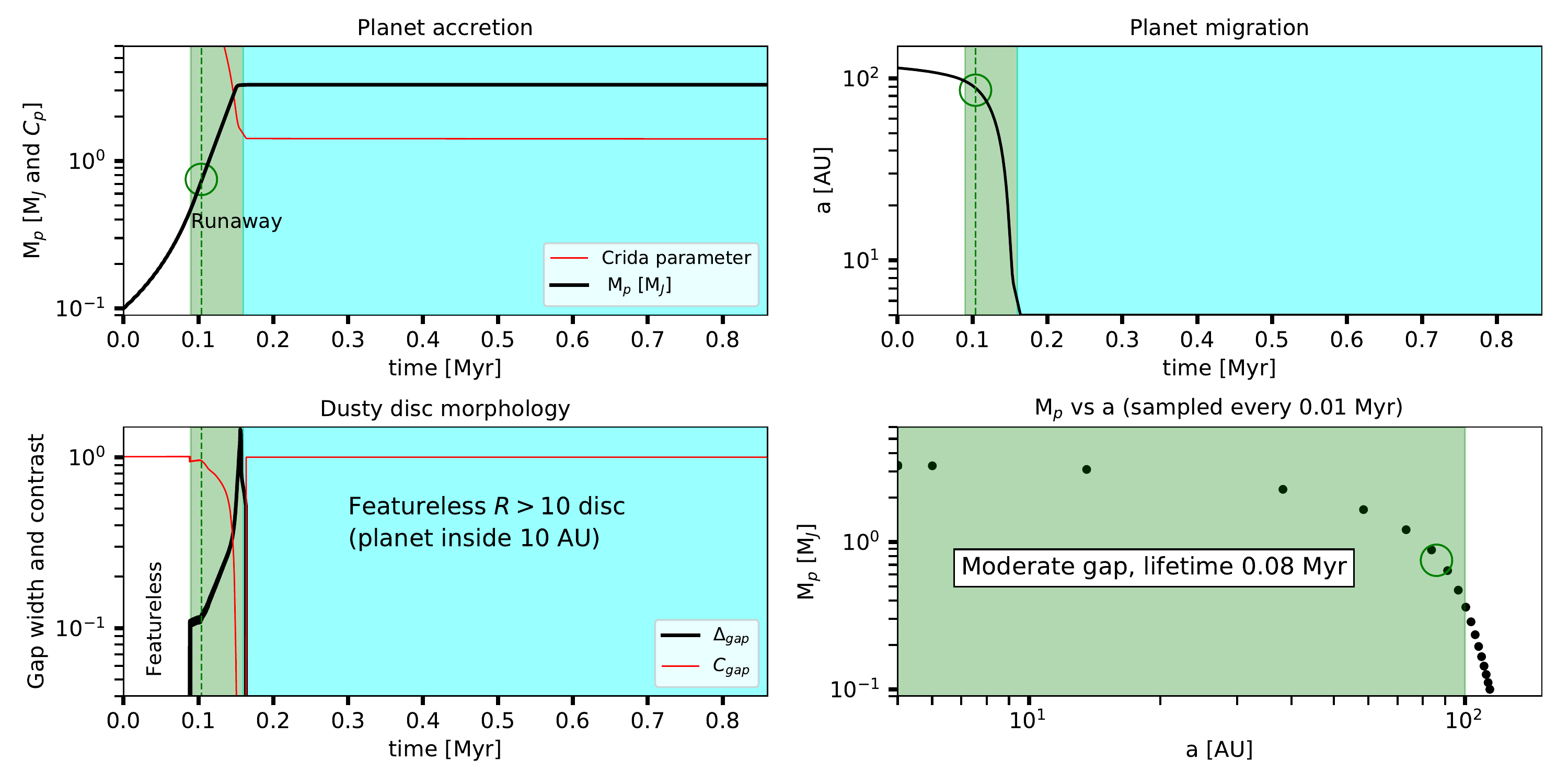}
\caption{Same as Fig. \ref{fig:HD-one-run} and Fig. \ref{fig:HD-alpha1e4} but for $\alpha_{\rm v} = 10^{-2}$. In this high viscosity case planet migrates inward too rapidly and does not open a gap until it is "too late", when it reaches the inner 10 AU disc. \bdf{This is the reason why there is no purple coloured background in this figure. Instead we use the cyan background to emphasise the times when there is no planet in the outer $R > 10$~AU disc since it migrated inside 10 AU. The outer disc will then look featureless in ALMA observations. The duration of the moderate gap case is again short, $0.08$~Myr.}}
\label{fig:HD-alpha1e2}
\end{figure*}

Experiments presented in \S \ref{sec:ss_start} are initialised assuming a non-migrating and non accreting planet. In a realistic situation the planet spends a long time in the relatively long low mass pre-runaway phase before embarking on the rapid mass gain \citep[e.g.,][]{PollackEtal96,MordasiniEtal12}. In this section we start instead with planets at the initial mass of $M_{\rm p} =0.1\mj$ which is high enough for the planet to enter the runaway phase, especially at large distances \citep[e.g.,][]{PisoYoudin14,Piso15}. We do not model how the planet grows to the initial mass as this is model dependent \citep[e.g., planetesimal vs pebble accretion, see][]{Bern20-1,NduguEtal19}. This important but model dependent issue is fortunately mainly irrelevant to our paper since very low mass planets do not produce significant gaps in dusty discs. This is certainly true for HD163296 as its star is one of the most massive and bright in \cite{LodatoEtal19} sample. Furthermore, below when we discuss the rest of the sample (\S \ref{sec:ALMA_sample})
we exclude from it all low mass planet candidates that may not yet be in the runaway regime.

Our approach in this section is to treat the location of the planet at the time it reached the runaway mass of $0.1\mj$ as a free parameter. We keep the same stellar parameters, the disc setup, and the maximum grain size as in \S \ref{sec:ss_start}.
By trial and error we found that planets starting at $a=114$~AU generally reach the desired mass, $M_{\rm p}\sim 0.5-1 \mj$, at around the gap location, and hence we present here only such experiments. We now present the simulations for three values of the disc viscosity parameter: $\alpha_{\rm v} = 10^{-2}, 10^{-3}$ and $10^{-4}$.

In order to relate to the observed parameteters of the $R=86$~AU gap in HD163296, we define the gap width, $\Delta_{\rm gap}$ as 
\begin{equation}
    \Delta_{\rm gap} = \frac{\Delta R_+ + \Delta R_-}{a}\;,
    \label{Delta_gap0}
\end{equation}
where $\Delta R_+$ is the distance between the planet location ($a$) and the centre of the ``ring" behind the planet. The ring position is defined as the farthest local maximum in $\Sigma_{\rm dust}$ (this definition avoids local peaks within the dusty gap which may be present for low values of $\alpha_{\rm v}$). $\Delta R_{-}$ is defined as the radial extent of the region inward of the planet location where the dust surface density is 10 times lower than the unperturbed dust density. We also define the gap contrast simply as the ratio
\begin{equation}
    C_{\rm gap} = \frac{\Sigma_{\rm min}}{\Sigma_{\rm max}}\;,
    \label{Cgap}
\end{equation}
where $\Sigma_{\rm min}$ and $\Sigma_{\rm max}$ are the dust minimum and maximum surface densities {\em over the gap region}. $\Sigma_{\rm max}$ in practice is usually found at the centre of the ring behind the planet (e.g., see the bottom left panel of fig. \ref{fig:HD163296}) or at the planet location for a rapidly migrating planet (e.g., the cyan curve in the top right panel of fig. \ref{fig:HD-one-run}). This definition is qualitatively similar to that used by \cite{RosottiEtal16} if the disc is optically thin and dust temperature does not vary appreciably over the gap region. We note that there is a degree of arbitrariness in these definitions, however they are conservative, and we found them to be more robust when the planet is allowed to migrate than definitions introduced by other authors in the past for static planets.

\subsubsection{Viscosity parameter $\alpha_{\rm v} =10^{-3}$.}\label{sec:HD_alpha1e3}

Fig. \ref{fig:HD-one-run} presents a calculation in which $\alpha_{\rm v} = 10^{-3}$. The simulation was terminated at $t \approx 0.85$~Myr when the planet reached 10 AU, twice the inner boundary of our computational domain. The top row of panels show several snapshots of the disc gas and dust surface density profiles,  similar to fig. \ref{fig:HD163296}. The middle and the bottom rows of panels detail the system evolution. The middle left panel shows planet mass and \cite{CridaEtal06} parameter $C_{\rm p}$, while the middle right shows planet separation, $a$, versus time. The bottom row shows the dust gap width and contrast versus time (left panel) and the planet mass versus position (right panel). The two bottom rows are shaded with three different colours to emphasise the three distinct states of the planet-disc system. 

Further, to expose the pace of planet evolution at different times, the planet mass-position diagram (bottom right panel) is sampled every 0.01 Myr. We see that at the beginning of the run, the planet gains mass relatively slowly as its ``feeding zone" is small, and its migration is also relatively slow. However, this accelerates strongly when the planet grows in mass by a factor of a few. The planet crosses the radius of 86 AU (the centroid of the dusty gap we study here) at $t= 0.104$ Myr. The planet mass at that moment is $M_{\rm p} = 0.67\mj$. While this mass is smaller than $1\mj$ we favoured in \S \ref{sec:ss_start}, we note that both the gap width and the gap contrast (the bottom left panel) are not far off from those inferred from observations of this gap \citep[$\Delta_{\rm gap} = 0.17$ and $C_{\rm gap} = 0.15$;][]{Dsharp7}. We place the green open circle on Fig. \ref{fig:HD-one-run} to mark that moment.

Prior to $t\approx 0.07$~Myr the gap depth is too small (that is, the gap contrast is $C_{\rm gap} \gtrsim 0.8$) to be compatible with the observations. Since such a gentle gap may  be unresolved (missed) in a disc less bright and not so well studied as HD163296, we term this early time period ``featureless"; \bdf{it has a white background} in the figure. During the time marked with the green background the parameters of the dusty gap are moderate, and are roughly consistent with that seen in HD163296. After $t\approx 0.15$~Myr the planet opens a gap too deep to be compatible with observations. During the phase that comes after $t\approx 0.15$~Myr, \bdf{coloured with purple background,} the disc morphology is that of a wide dust and gas gap transitioning into a complete inner hole in dust (cf. the green dotted curve in the top right panel).

Note that the green region on the plot corresponds approximately to the runaway phase of the planet growth, whereas the wide gap/inner hole disc state corresponds quite well to the detached phase when the planet stops growing. This could also be rephrased in terms of the two planet migration regimes. The wide gap/inner hole disc state corresponds to the planets migrating in type II regime, when $C_{\rm p} \leq 1$ (compare the left panels in the middle and the bottom rows). This is unsurprising since planets open deep gaps in the {\em gas} disc when $C_{\rm p} \leq 1$ \citep{CridaEtal06}; the gaps and holes in the dust discs are always stronger than those in the gas discs \citep[e.g.,][]{DipierroEtal16a,DipierroLaibe17}. This correspondence between the dusty disc morphology, on the one hand, and the planetary growth and migration phases, on the other, is exciting: ALMA may be able to probe planetary growth and migration in the CA scenario directly by the statistics of gaps/holes in protoplanetary discs.

Quantitatively, the main conclusion from this numerical experiment confirms what we found from the steady state start models in \S \ref{sec:ss_start}: the planet-disc system spends a surprisingly short  period of time in the corner of the parameter space consistent with observations. The duration of this phase is no longer than 0.07 Myr, which is comparable with the results we obtained from the steady state start models.

\subsubsection{Viscosity parameter $\alpha_{\rm v} =10^{-4}$.}\label{sec:HD_alpha1e4}

Fig. \ref{fig:HD-alpha1e4} shows a numerical experiment fully analogous to that presented in Fig. \ref{fig:HD-one-run}, \S \ref{sec:HD_alpha1e3}, but with a much lower viscosity parameter, $\alpha_{\rm v} = 10^{-4}$. The initial evolution of the planet is very similar to that in fig. \ref{fig:HD-one-run}. However, due to lower viscosity, a gap in the disc is opened earlier, and the planet switches to type II migration upon reaching the radius of 50 AU, rather than $\sim 30$~AU. The effects of the planet on the dust also become noticeable much earlier, so that the featureless disc phase terminates even sooner. 

The largest difference with the $\alpha_{\rm v} = 10^{-3}$ case is however the much slower rate of planet migration in the Type II regime. Indeed, during the 1 Myr of this simulation the planet migrates only to about 33 AU. This has major observational implications for the appearance of the dusty gap. The gap in this case is situated farther out, and is much wider and deeper. \bdf{Once again, the gap contrast is formally zero after $t\sim 0.15$~Myr, which clearly rules this scenario out as a reasonable model for the gap we study here. }

\subsubsection{Viscosity parameter $\alpha_{\rm v} =10^{-2}$.}

Fig. \ref{fig:HD-alpha1e2} shows yet again the same numerical experiment but with a higher value of the viscosity parameter, $\alpha_{\rm v} = 10^{-2}$. In this case we observe major qualitative differences in the disc-planet system evolution. The planet does not manage to open a deep gap in the disc ($C_{\rm p} > 1$) until it reaches the inner boundary of our computational domain, exiting the grid.  Except for a very brief time, the gap contrast and the gap width are never large enough to classify this system as a wide gap system. \bdf{This is why there is no region with purple background in Fig. \ref{fig:HD-alpha1e2}. } After the planet exits the computational grid through the inner boundary at 5 AU, the outer disc looks featureless. \bdf{We use cyan background to highlight this state of the disc-planet system.} Note that our results would not be much different even if we extended the grid to much smaller radii. Indeed, the type II migration in this case is quite rapid and so the planet would migrate "into the star" quickly. Even if the planet were to hang about in the inner disc for an unspecified reason its torques would not affect the disc at radii that ALMA can resolve in a typical source. \bdf{Finally, molecular line observations constrain $\alpha_{\rm v}\lesssim 3\times 10^{-3}$ \citep{Flaherty17-HD163296-turb}. Concluding,} this high viscosity parameter experiment is also not promising as an explanation for the ALMA gap at 86 AU in HD163296.

\subsection{Observational implications for the Fiducial model of HD163296 gap at 86 AU}

None of the three viscosity values considered above yielded long-lived moderate gap discs, although for different reasons. To summarise, for the lowest viscosity parameter, the planet runs away in mass to become a massive gas giant at wide separations (tens of AU). It opens a deep gap/inner hole and migrates inward very slowly in the type II regime. At the highest $\alpha_{\rm v}$ explored, the planet does not fully reach its gas gap opening mass before it reaches the inner $\sim 10$~AU disc, so it has a moderate gap while at wide separations. However, the planet migrates from 100 AU to 10 AU in $0.08$~Myr. The intermediate viscosity case, $\alpha_{\rm v} =10^{-3}$, shows elements of both of these extreme cases, but the end result is nonetheless the same: the moderate gap state lasts time $\Delta t_{\rm mod}\sim 0.08$~Myr. 

\begin{figure*}
\includegraphics[width=0.99\textwidth]{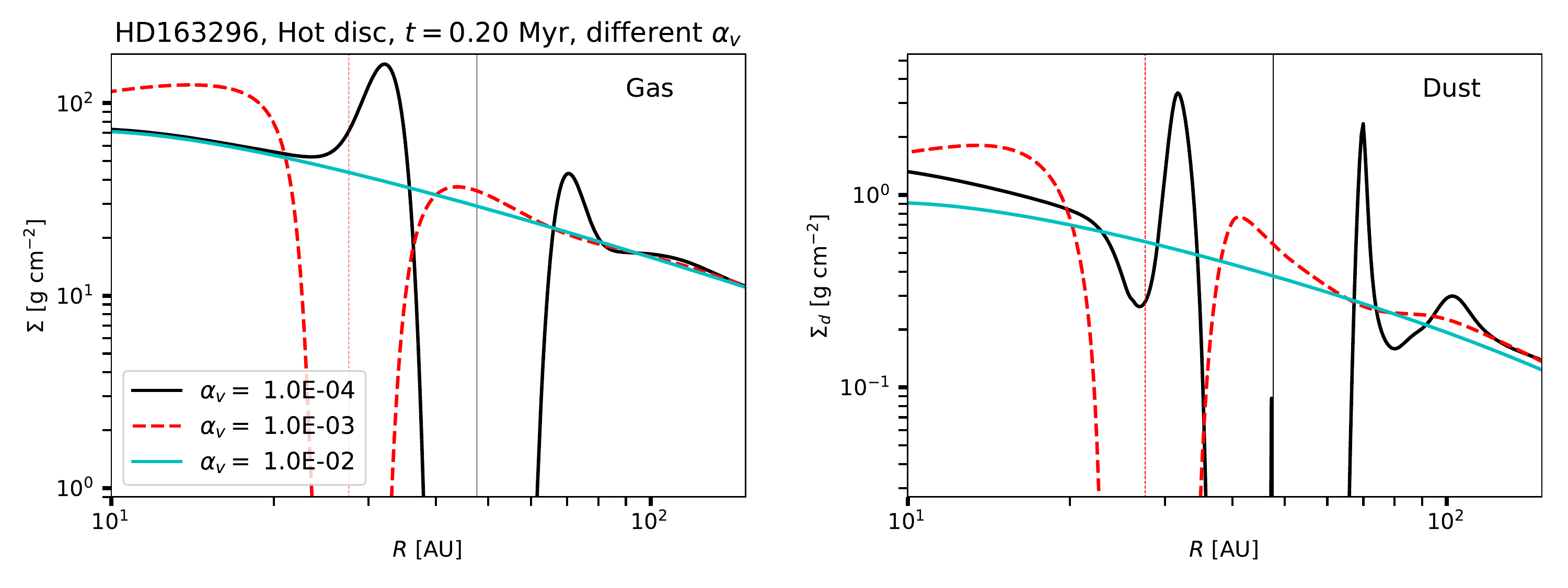}
\caption{The gas and dust surface density profiles at time $t=0.2$~Myr for the fiducial simulations with the three values of the viscosity parameter. This shows that only $t=0.2$ Myr after the beginning of the runaway accretion the dusty disc morphology is not right compared with the observed one.}
\label{fig:Disc_at_same_t}
\end{figure*}

A corollary of this rapid planet evolution is the fact that after just a brief period of time the disc in HD163296 would look very different from the observed one. Fig. \ref{fig:Disc_at_same_t} shows the gas and dust surface density profiles at $t=0.2$~Myr for these three runs. At the lowest viscosity parameter, the gap width and depth are too large compared with the observed values; at the highest $\alpha_{\rm v}$ the planet is lost into the inner disc and so the outer disc is featureless in both gas and dust; and at $\alpha_{\rm v} = 10^{-3}$ the planet both grew and migrated in too much. Furthermore, these conclusions are robust to changes in the assumed temperature profile as shown in Appendix \ref{sec:app_numerics}.

The problem of too rapid planet evolution can also be understood via the following argument. Let us consider an ensemble  of protostars with stellar and disc parameters as we assumed for HD163296, and let planets grow in these discs according to the CA runaway scenario. We assume that the ages of these systems is a uniformly distributed variable between 0 and $\sim 2t_0 \sim 10$ Myr. If each system hatches one embryo massive enough to get into the runaway regime, then we have a chance $\Delta t_{\rm mod}/2t_* \sim 0.01 $ to catch any given system in the moderate gap state. If we pick one system from this ensemble in random, it would be very surprising if it were a disc with a moderate gap. 
Further, if we observed a number of randomly selected objects in this ensemble then we should expect that most of them would be in either featureless (white background in Figs. \ref{fig:HD-one-run}, \ref{fig:HD-alpha1e4}, and the cyan one in \ref{fig:HD-alpha1e2}) or the deep gap/inner hole, the low viscosity detached planet case (the purple background in the figures). The featureless state corresponds to the times before a massive core has grown and/or the planet migrated inside 10 AU. The wide gap/inner hole state is realised at low viscosity parameter values when the planet enters the type II regime.

This predicts a lack of systems with moderate gaps, and abundance of featureless and wide gap discs. These theoretical predictions are rather divergent from what we see for the observed sample of ALMA discs/planets \citep{LodatoEtal19}. 
 
\subsection{Suppressing gas accretion and migration}\label{sec:Fa10}

\cite{NayakshinEtal19} showed that decreasing the gas accretion rate onto planets while they are on the runaway track of CA by $F_{\rm acc} \sim$ an order of magnitude brings CA theory predictions much more in line with the mass function of candidate ALMA planets. This finding is also consistent with a number of recent simulations of gas accretion onto CA planets which suggests that gas accretion may be much less efficient in the runaway phase than previously believed \citep[e.g.,][]{Szulagyi14,OrmelEtal15,FungChiang16,LambrechtsLega17}. 

We therefore investigate how suppressing gas accretion by a factor $F_{\rm acc} =10$ would affect our results.  \cite{NayakshinEtal19} held their planets at fixed orbits artificially whereas our model includes planet migration. Experiments presented in Fig. \ref{fig:HD163296} demonstrated that the planets held at a constant mass (which is equivalent to $F_{\rm acc} = \infty$) are still challenged by the ALMA data as planet migration may also be too rapid. Therefore we also introduce a factor $F_{\rm mig} \geq 1$ to reduce the rate of planet migration. Similar approach was chosen in the past by many authors, e.g., \cite{IdaLin04b}.

\begin{figure}
\includegraphics[width=0.49\textwidth]{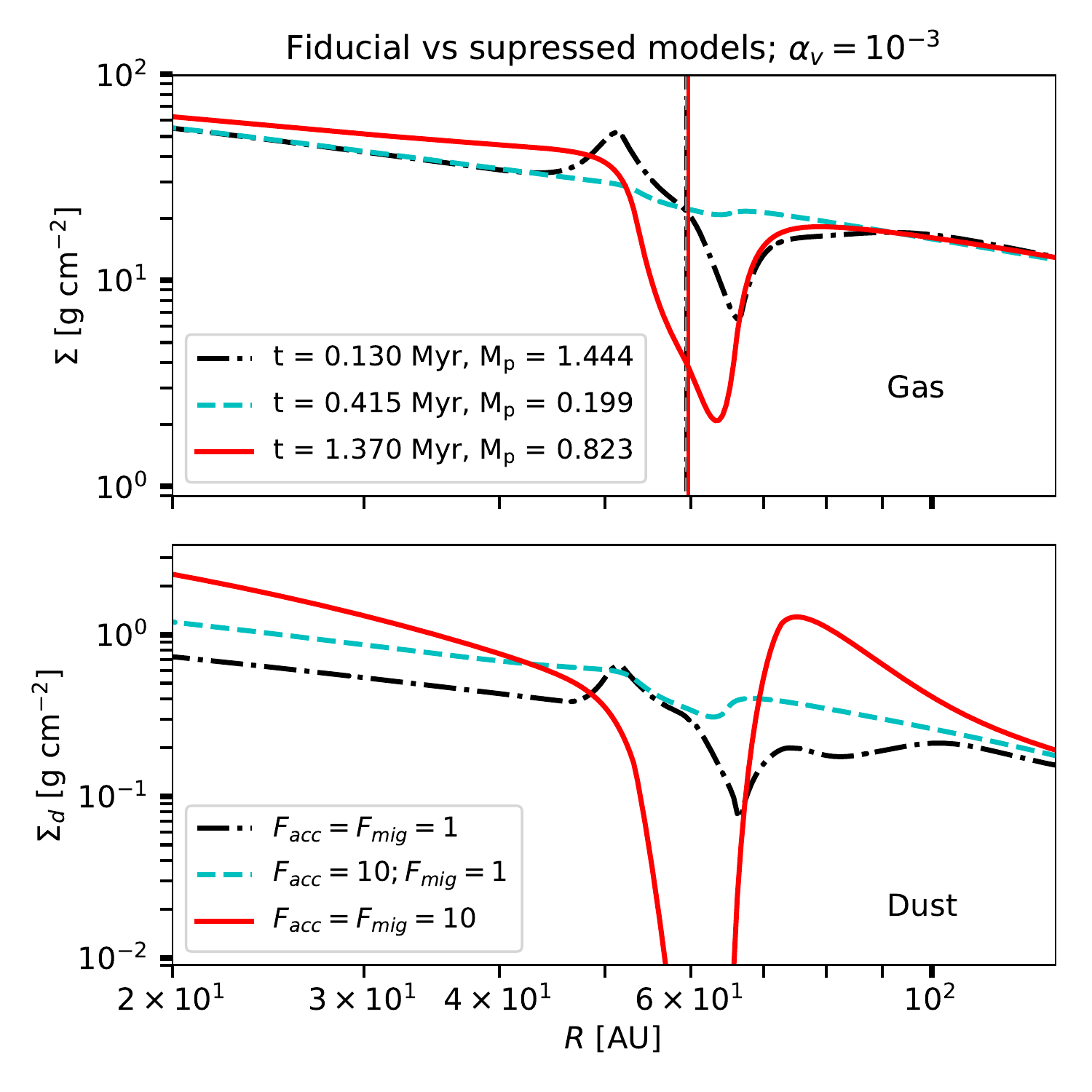}
\caption{Comparison of the disc surface density profiles of the nominally growing planet, previously shown in fig. \ref{fig:HD-one-run}, with that of suppressed evolution simulations. It is now plainly obvious that the dust surface density in the fiducial simulation (black dot-dashed curves) is not an appropriate model to account for ALMA observations because the bright ring is in front of the planet, not behind. If only planet accretion is suppressed (cyan) then the planet never grows massive enough to affect the disc significantly enough before it migrates into the inner disc.  The dust profile with both accretion and migration suppressed, on the other hand, is qualitatively reasonable, and is much longer lasting.}
\label{fig:HD-Facc-Fmig-10}
\end{figure}

Fig. \ref{fig:HD-Facc-Fmig-10} compares the gas and dust surface density profiles of the disc in the fiducial $\alpha_{\rm v} = 10^{-3}$ simulation presented in Fig. \ref{fig:HD-one-run} with that performed with slow-down factors $F_{\rm acc} = 10$ and $F_{\rm mig} = 1$ (slow accretion but nominal migration), and $F_{\rm acc} = F_{\rm mig} = 10$ (both accretion and migration suppressed). As in the fiducial case, the planet starts with mass $M_{\rm p} = 0.1\mj$ and the initial separation of $a_{\rm p} = 114$~AU. In Fig. \ref{fig:HD-Facc-Fmig-10} we compare the disc surface density profiles when the planet positions coincide, at $a\approx 60$~AU. This occurs at different times in the simulations, and at different planet masses. We note that our main conclusions do not depend on the  planet position chosen for this illustrative comparison.

In the suppressed accretion but fiducial migration scenario (cyan curves in the figure) the planet migrates inward so rapidly that it does not manage to grow enough to affect the disc sufficiently. Its mass is only $0.2\mj$ and with the chosen $\alpha_{\rm v}$; this is plainly too low to open a detectable gas. This scenario looses planets into the inner disc too rapidly to be compatible with observations of HD163296.

In the case when both accretion and migration are suppressed (red solid curves), we see that the planet is much more massive by the time it reaches $a=60$~AU. The planet mass is lower in the suppressed simulation than in the fiducial simulation, but the gap opened by the planet is deeper in both gas and dust. Additionally, in the fiducial simulation, there is a significant radial offset between the positions of the gap and the planet. This is because a very rapidly migrating planet may be said to push the material in front of it, sort of snow ploughing it,  creating the dusty ring in front of itself rather than behind it \citep{Meru19-Ring}. For the ring to be behind the planet, the planet cannot migrate inward much more rapidly than the dust drifts in. HD163295 is $\approx 5$~Myr old. Therefore a planet migrating on a time scale of $\sim 0.1$~Myr could not possibly be providing a barrier to hold the dust back. If this were the rule for protoplanetary discs then it would be hard to invoke planets as the mechanism for slowing down dust radial drift \citep[as argued in, e.g.,][]{PinillaEtal12}, leaving additionally the conundrum of dust survival in old discs unsolved \citep[e.g.,][]{Birnstiel09,BirnstielEtal12}. 

We also note the general success of the stationary (thus $F{\rm mig} \gg 1$) planet models employed by \cite{Dsharp7} to explain the DSHARP sample of ALMA gaps/rings. If the planets migrated rapidly in as predicted by the standard theory, then the planetary scenario for the origin of gaps would not be so convincing. Slowly migrating planets, such as the simulation with $F_{\rm mig} =10$ in fig. \ref{fig:HD-Facc-Fmig-10} are clearly more promising in addressing this issue.

The fiducial model also probably contradicts observations of the gas kinematics in HD163296 as well. \cite{Pinte20-Dsharp-Vkinks} show that the localized velocity perturbation (kink) found in HD163296 near the gap at 86 AU falls right into the middle of the gap after the deprojection of the CO emission map. That is, the radial location of the planet coincides with the centroid of the dusty gap. The fiducial model, however, shows a significant offset between the location of the planet and the dusty gap position. The suppressed migration and accretion model does not have such an offset and are thus more promising.

\begin{figure*}
\includegraphics[width=0.99\textwidth]{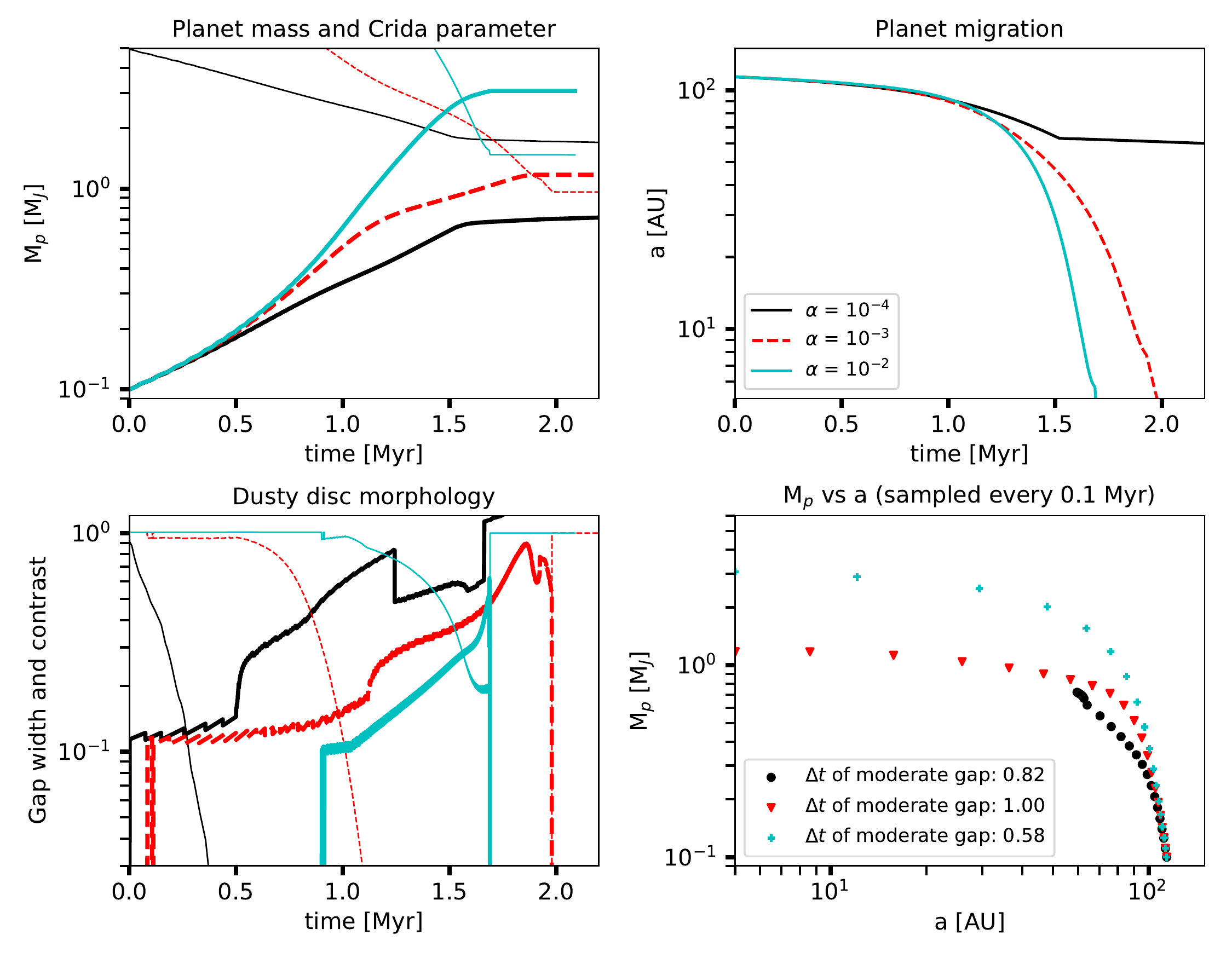}
\caption{Same as figs. \ref{fig:HD-one-run} to \ref{fig:HD-alpha1e2} but for the simulations in which both migration and accretion are suppressed by one order of magnitude, $F_{\rm acc} = F_{\rm acc} = 10$ (see \S \ref{sec:Fa10}) and for three different values of the disc viscosity parameter $\alpha_{\rm v}$ as listed in the legend in the top right panel. The times spent by the disc in the moderate gap configuration are shown in the bottom right panel in the legend. Note that these are much longer than those of fiducial simulations shown in Figs. \ref{fig:HD-one-run} -- \ref{fig:HD-alpha1e2}.}
\label{fig:HD-History-F10}
\end{figure*}

Fig. \ref{fig:HD-History-F10} shows the time evolution of the planet-disc system for the suppressed planet migration and accretion scenario, $F_{\rm acc} = F_{\rm mig} = 10$, now for three values of the viscosity parameter: $\alpha_{\rm v} = 10^{-4}$, $10^{-3}$ and $10^{-2}$. The meaning of all of the curves is the same as in fig. \ref{fig:HD-one-run}. While qualitatively this evolution may look similar to that shown in figs. \ref{fig:HD-one-run} to \ref{fig:HD-alpha1e2}, the discs spend much more time in the state of moderate gaps, here defined as $C_{\rm gap} \leq 0.9$ and $\Delta_{\rm gap} \leq 0.4$. These times are listed in the caption in the bottom right panel, and vary between 0.6 and 1 Myr. This is much more reasonable. If a few massive planet embryos formed in the disc of HD163296 in the last few Myr, then we would have a decent statistical chance to detect at least one of them in the state of moderate gap now.

\section{Results for the ALMA gap sample}\label{sec:ALMA_sample}

We now use the  sample of ALMA gaps/candidate planets investigated in \cite{NayakshinEtal19} \citep[the wast majority of which is from][]{LodatoEtal19} in order to see how the lifetime of the observed gaps could constrain the CA scenario for planet formation. To do this, we first remove from the sample any gaps that are believed to be opened by planets less massive than the mass for which the planet should be in the runaway CA regime, which we define here as $0.1\mj$. We also chose only the gaps located beyond 10 AU. We have 21 gaps remaining in the sample.  The parameters of these systems are presented in Table 1.

We then follow the ``release from the steady state" approach from \S \ref{sec:ss_start} rather than the ``release at the runaway" approach from \S \ref{sec:runaway_start}. In the latter, the initial location of the embryo is unknown and thus one needs to make guesses and iterate to constrain it. The number of runs required to match our sample of ALMA gaps and to then also study the implications for the CA scenario is too large for such trial-and-error investigation. On the other hand, in \S \ref{sec:HD_SS} we found that the simpler ``release from the steady state" approach yields quantitatively similar value for the duration that the gaps may remain in the observed state.

For each object in the sample we set up a steady-state power-law disc model with the same radial extent as before (5 to 200 AU). We determine the disc mass via the following constraints. First of all, given the observed accretion rate onto the star, $\dot M_*$, and its age, $t_*$, we require that there must be at least $M_{\rm acc} = 2 \dot M_* t_*$ of gas to maintain this accretion rate \citep[e.g.,][]{JPA12}. Note that this minimum disc mass estimate simply states that disc is the source of gas accretion onto the star and must have fed the star for time $t=t_*$. The factor of 2 in the estimate can be larger (up to $\sim 10$) and accounts for the fact that matter also spreads outward \citep{JPA12}. This estimate is conservative and is likely to hold no matter the detail of the exact mass transfer mechanism from the disc to the star. For example, the recent MHD disc wind formalism developed by \cite{Tabone22-general} will increase the disc mass estimate because only a fraction of the disc mass reaches the star in this case.

Next, following the same arguments as \cite{Powell17-dust-lines,Powell19-DustLanes} we require the drift timescale for 1 mm sized grains at the location of the planet to be at least as long as $t_*$. While planets are often invoked as a solution to the short dust radial drift times conundrum \citep[e.g.,][]{PinillaEtal12}, most of the observed ALMA gaps are quite shallow and hence may at best slow down but not prevent the drift. We term this minimum disc mass estimate as $M_{\rm drift}$.

We then take the disc mass to be a maximum of $M_{\rm acc}$ and $M_{\rm drift}$. However, \cite{HallEtal18} show that discs that exceed mass $M_{\rm sg} = 0.25 M_*$ become very strongly self-gravitating. Spiral density arms in such discs are likely to be detectable with ALMA, and transfer the mass onto the star at very high rates, depleting the discs of mass in just $\sim 10^4$ years. Both of these effects strongly contradict what is observed for the ALMA sample: the discs have annular symmetry and accretion rates onto the star some $\sim 2-4$ orders of magnitude smaller. We therefore limit the disc mass by $M_{\rm max} = 0.2 M_*$. 

The disc surface temperature profile is set in this modelling by
\begin{equation}
    T(R) = \left[\phi \frac{L_*}{4\pi \sigma_B R^2} \right] ^{1/4}\;,
    \label{TofR}
\end{equation}
where $\phi = 0.1$ and $L_*$ is the luminosity of the star \citep[this model is similar to that used by, e.g.,][]{LongEtal18}. We set the disc viscosity parameter to $\alpha_{\rm v} = 10^{-3}$ in this section.

We then insert a planet with the mass inferred from the gap properties at the location of the observed gap into the unperturbed disc. Planet growth and migration are not allowed for 0.5 Myr to allow the disc-planet system to reach a steady state. We then  allow the planet to gain mass by accretion and migrate. We define the planet migration and planet accretion time scales as
\begin{equation}
    t_{\rm mig} = - \frac{a}{\dot a}\;,
    \label{tmig_def}
\end{equation}
and 
\begin{equation}
    t_{\rm acc} = \frac{M_{\rm p}}{dM_{\rm p}/dt}
    \label{tacc_def}
\end{equation}
respectively. Given the close association between the gap properties and the planet mass and separation we define the gap lifetime as the following: 
\begin{equation}
    t_{\rm gap}^{-1} = t_{\rm acc}^{-1} + t_{\rm mig}^{-1}
\end{equation}
Note that this definition is a smoothed out alternative to $t_{\rm gap} = \min[t_{\rm acc}, t_{\rm mig}]$.

\begin{figure}
\includegraphics[width=0.49\textwidth]{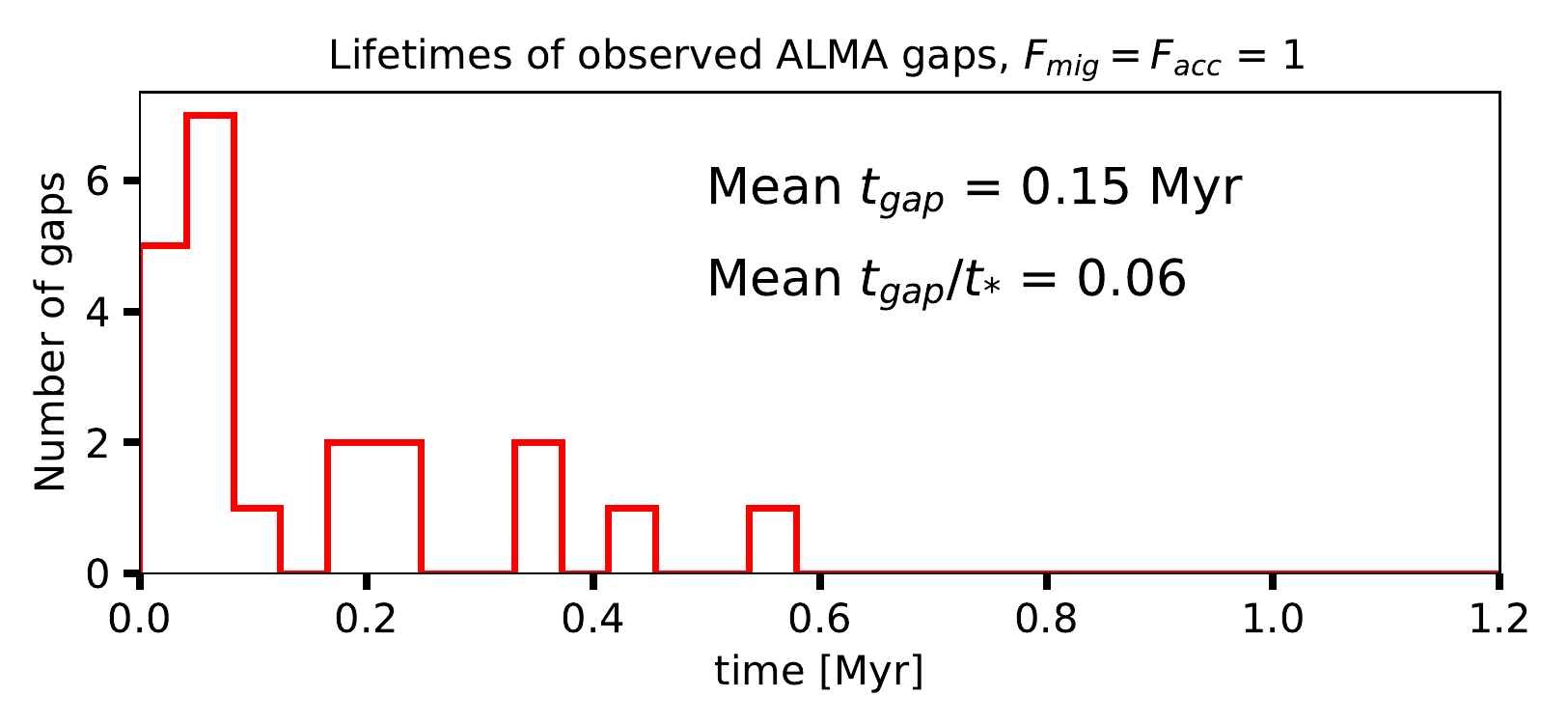}
\includegraphics[width=0.49\textwidth]{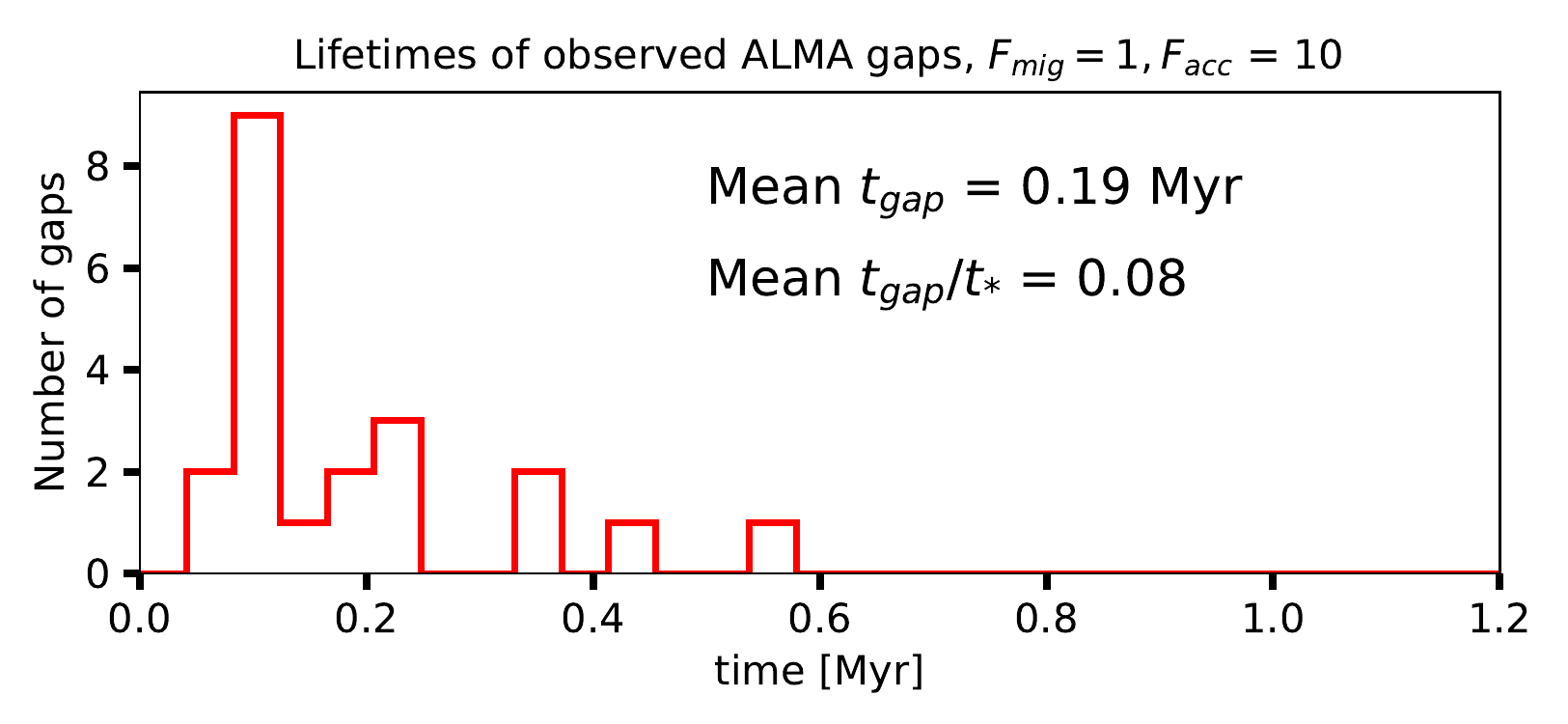}
\includegraphics[width=0.49\textwidth]{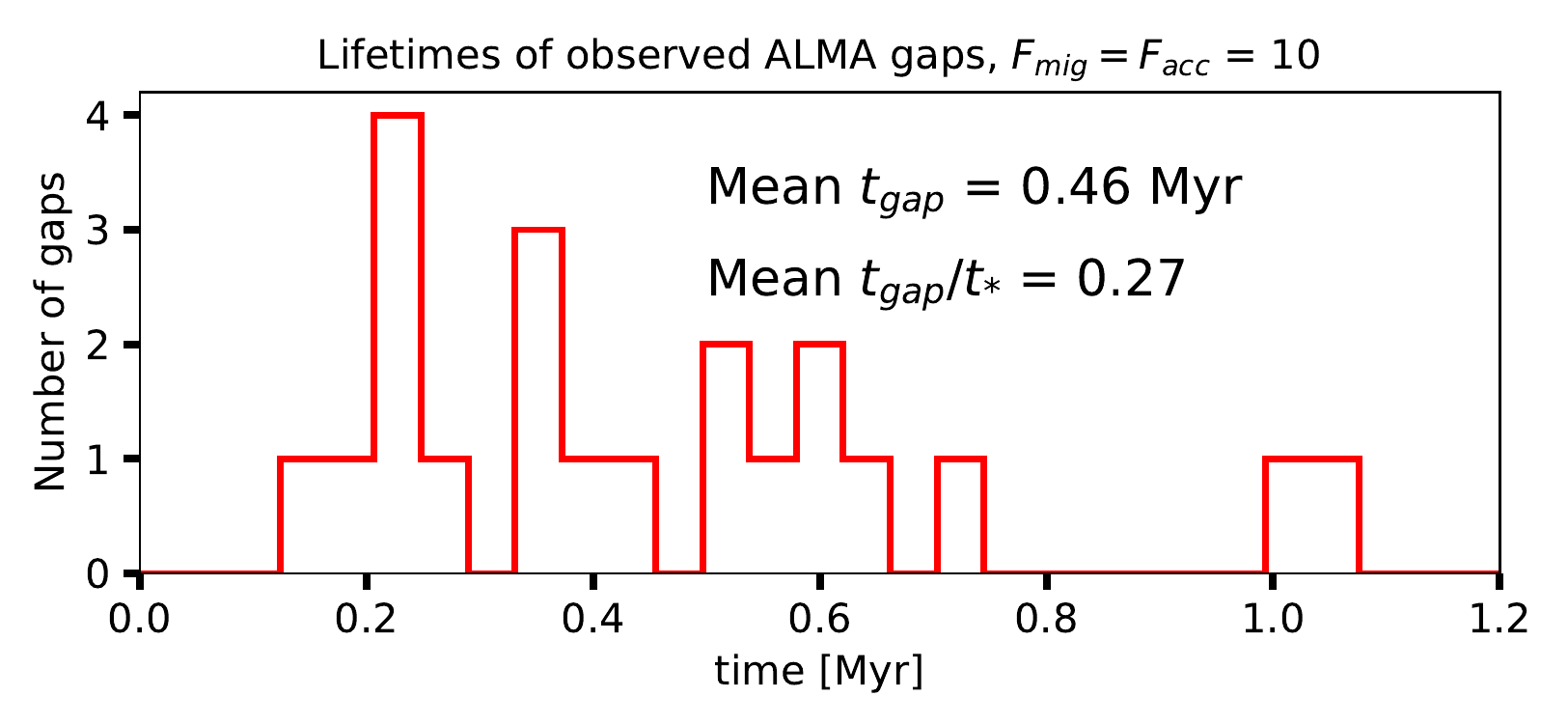}
\caption{Distribution of the observed gap lifetimes from the ALMA sample for fiducial ({\bf top panel}), accretion suppressed ({\bf middle panel}), and both accretion and migration suppressed scenarios ({\bf bottom}). Only the last of the three cases provides reasonably long gap lifetimes.}
\label{fig:Lodato19_sample_dt_histograms}
\end{figure}

The top panel in Fig. \ref{fig:Lodato19_sample_dt_histograms} shows the histogram of $t_{\rm gap}$ evaluated with the fiducial CA model for our 21 gaps in the sample. Table 1 shows $t_{\rm gap}$ for each individual gap/planet. We observe in Fig. \ref{fig:Lodato19_sample_dt_histograms} that for most of the gaps  $t_{\rm gap} < 0.1$ Myr, although there are some systems with longer $t_{\rm gap}$. The mean value of $t_{\rm gap}$ for the sample is $0.15$ Myr. These time scales are worryingly short compared with the system ages, $t_*$, which are in the range of 1-10 Myr. More specifically, the mean of the ratio $t_{\rm gap}/t_*$ is $0.06$ for the sample. 

In the middle panel of Fig. \ref{fig:Lodato19_sample_dt_histograms} we test the suppressed gas accretion scenario, setting $F_{\rm acc} = 10$, but allowing planets to migrate at the fiducial rate ($F_{\rm mig}=1$). This extends the gap lifetime somewhat, especially for the shortest $t_{\rm gap}$ bins. However, $t_{\rm gap}$ is still unacceptably short on average. 

Finally, in the bottom panel we investigate the scenario in which both gas accretion and planet migration are suppressed by a factor of 10. In this case, the typical gap lifetime is $\sim 0.5$ Myr, and the mean for the ratio $t_{\rm gap}/t_* = 0.27$. These results are much more reasonable, requiring a few gas giant planets to be born per system in this sample.

\begin{figure}
\includegraphics[width=0.49\textwidth]{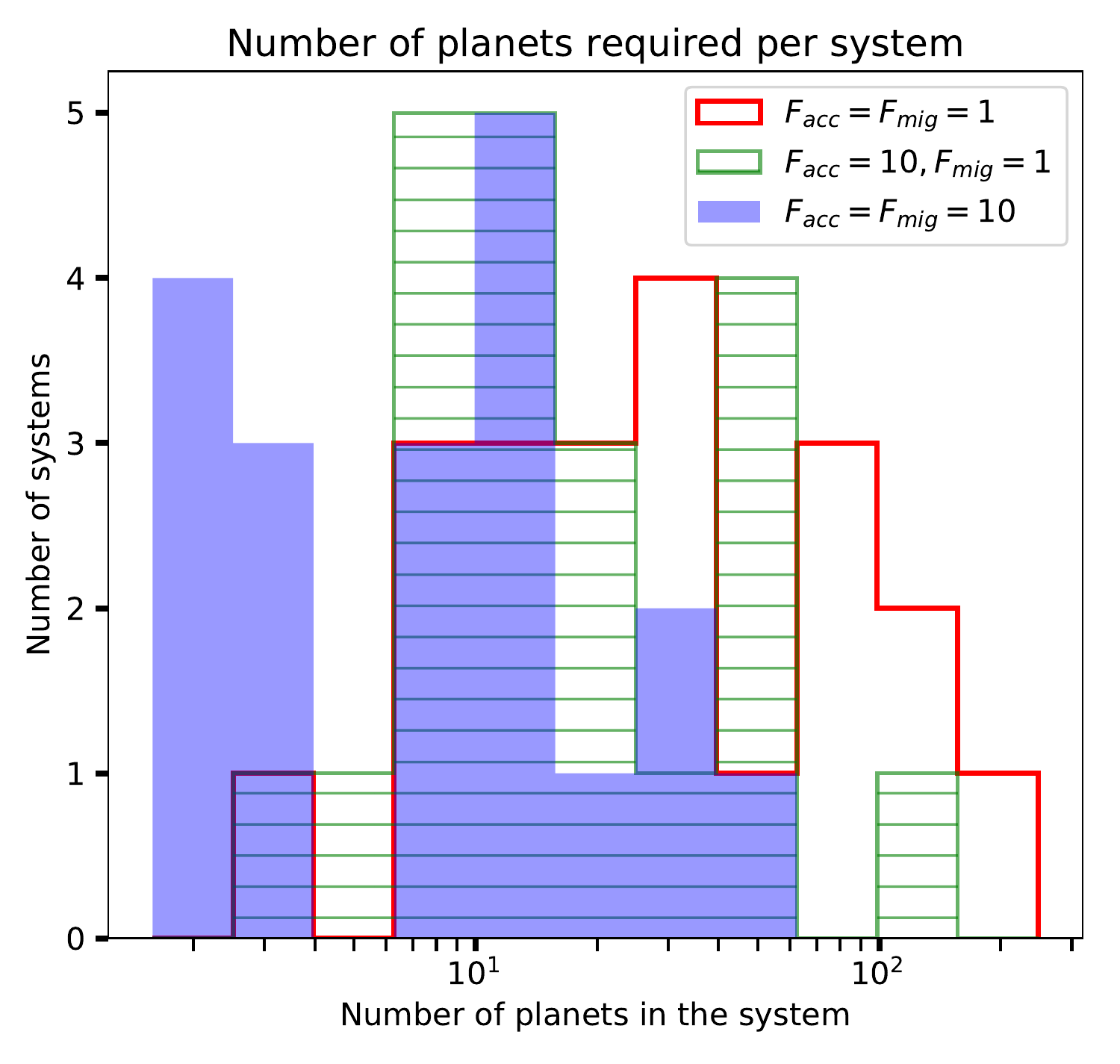}
\caption{Number of planets per system required for a probability of order unity for ALMA to detect a planet in that system (cf. eq. \ref{N_req0}).}
\label{fig:Lodato19_N_required}
\end{figure}

In Fig. \ref{fig:Lodato19_N_required} we show the number of planets that has to be born in the system to yield a probability of order unity for us to observe it,
\begin{equation}
    N_{\rm req} \sim \frac{t_*}{t_{\rm gap}}
    \label{N_req0}\;.
\end{equation}
The red open histogram shows this quantity for the fiducial model, that is, for $F_{\rm acc} = F_{\rm mig} = 1$. In this scenario, for gaps opened by planets to be as common as implied by the observations we would need tens to a hundred of gas giant planets to be born per star, which is clearly excessive. For slower gas accretion onto planets, $F_{\rm acc}=10$, we obtain the green shaded histogram in Fig. \ref{fig:Lodato19_N_required}. This however relaxes the constraint only by a factor of $\sim 2$. Finally, the model where both accretion and migration are suppressed by an order of magnitude (the blue filled histogram) needs $\sim 10$ gas giant planets per star, although this is a significant dispersion around this number, with many systems requiring just a few planets whereas for some a few tens is needed.

The model with suppressed planet evolution, $F_{\rm acc} = F_{\rm mig} = 10$ is clearly favoured compared with the predictions of the other two scenarios. 
While $N_{\rm req}$ is still somewhat uncomfortably large for a few systems in the suppressed scenario, it is not yet clear if this constitutes a problem. It is possible that future better observations may reduce $N_{\rm req}$. We also need to remember that the present ALMA sample of resolved discs is dominated by the brightest systems \citep{Dsharp1,LongEtal18}, and hence they may be especially effective in spawning gas giant planets.

\section{Discussion}\label{sec:Discussion}

\subsection{Main results}

In this paper we \bdf{studied} 1D models of dusty protoplanetary discs with embedded gas giant planets growing according to the Core Accretion scenario. \bdf{We found  
a kinship between planet growth phases and the dusty disc morphology. As is well known \citep[e.g.,][]{PollackEtal96}, on the way to becoming gas giant planets, CA planets go through the three main phases: (i) a low mass planetary core; (ii) a moderate mass planet in the runaway gas accretion phase; (iii) a massive gas giant planet that stopped growing by opening a deep gas gap in the disc.  We find that in the pre-runaway phase (i), the planet is too low mass to open a detectable gap even in the dust disc. Barring other mechanisms creating dust rings/gaps in discs, phase (i) hence corresponds to featureless, smooth, protoplanetary discs. In the phase (ii) the planet mass is $\sim 0.1-0.5\mj$. Such a planet is massive enough to open gaps in the dust but not gas \citep[e.g.,][]{DipierroEtal16a,RosottiEtal16}. This phase is probably the one most frequently seen in ALMA observations of bright discs \citep[e.g.,][]{Dsharp2,Dsharp7,LongEtal18}. Finally, when the planet opens a deep gap in the gas disc, there is a very wide and deep gap in the dust disc, evolving into an inner disc hole if the mm-sized dust in the inner disc region is completely cutoff from the outer disc \citep{RiceEtal06}.}

\bdf{Significantly,} the theory predicts that \bdf{phases (i) and (iii) are long lived, whereas phase (ii) is very short. As expected based on classical works \citep[e.g.,][]{PollackEtal96,IdaLin04a}, we found that}  gas giant planets grow through the mass range $M_{\rm p}\sim 0.1\mj $ to a (gas) gap opening mass in a matter of $\lesssim 0.1$~Myr. \bdf{This implies that in the CA scenario the featureless discs, and the discs with very deep and wide holes, should be the mode of protoplanetary discs, whereas the moderate dust-only gaps should be a minority of discs transiting from phase (i) to phase (iii). We compared this prediction with} the statistics of the observed annular structures in the samples of resolved protoplanetary discs \citep{Dsharp1,LongEtal18,LodatoEtal19} \bdf{and uncovered} a significant mismatch between theoretical predictions and observations. \bdf{On the quantitative level,} the gap opening mass is a function of the poorly constrained disc viscosity parameter $\alpha_{\rm v}$; however, the predictions of the theory disagree with observations for any $\alpha_{\rm v}$.

At low values of $\alpha_{\rm v} \ll 10^{-3}$, planets open deep gaps in the disc quickly. Their growth therefore terminates at lower masses, typically $M_{\rm p} \lesssim 1\mj$. They open a deep gap in the gas disc, and migrate inward very slowly in the type II regime. This slow migration could explain the surprisingly large abundance of massive gas giant planets observed by ALMA; however, the  morphology of dusty discs in this case is deviant from what is observed. Planets that open deep gaps in gas discs filter out large dust particles very efficiently \citep[e.g.,][]{RiceEtal06},  leading to very wide gaps turning into complete inner holes in mm-sized dust discs (see fig. \ref{fig:HD-alpha1e4}). If this was the case then we would expect most ALMA discs look like wide gap transition discs, such as the famous PDS70 system where two massive gas giants opened an inner dust hole that extends from $\sim 2$~AU from the star to $\sim 30$~AU \citep[e.g.,][]{HaffertEtal19,KepplerEtal19,BenistyEtal21}. However, this system is unique amongst the ALMA systems in many ways, with most other discs showing the signs of mild gaps.

At high values of the disc viscosity parameter, $\alpha_{\rm v} \gg 10^{-3}$, on the other hand, planets are not able to open deep gaps in the gas disc until they reach higher masses, $M_{\rm p} \sim$ a few $\mj$ typically. They therefore migrate in the type I regime for longer, and are thus less likely to create discs with wide inner dust hole morphology. However, these planets migrate into the inner disc too rapidly, making it hard to understand the ubiquity of moderate gaps at large separations observed by ALMA (unless they are not opened by planets).

At intermediate values of viscosity parameter, $\alpha_{\rm v} \sim 10^{-3}$, models experience a mix of the two difficulties explained just above. Additionally, we found that for any value of disc viscosity the phase at which the planets produce moderate gaps in the disc, $t_{\rm gap}$, is too short, $t_{\rm gap} \lesssim 0.1$~Myr (the phases shaded green in figs. \ref{fig:HD-one-run} to \ref{fig:HD-alpha1e2}). We can estimate the probability of observing a phase of duration $t_{\rm gap}$ in a system with age $t_*$ as $t_{\rm gap}/t_*$. Given that most bright ALMA discs show dusty gaps, we require $N_{\rm req} \sim t_*/t_{\rm gap}$  gas giant planets born per ALMA disc. This results in $N_{\rm req}$ in many tens to a few hundred (cf. the red open histogram in fig. \ref{fig:Lodato19_N_required}), which is clearly excessive and would violate the available solid and (even) gas mass budget of protoplanetary discs.

In \S \ref{sec:Fa10} we therefore explored the  ``suppressed" planet evolution models. \bdf{Early one dimensional planet envelope contraction models \citep[e.g.,][]{Mizuno80,Stevenson82,IkomaEtal00} arrived at very high gas accretion rates in the runaway regime. However, 3D} simulations by various authors in the recent past \citep[e.g.,][]{Szulagyi14,OrmelEtal15,LambrechtsLega17} showed that gas accretion onto planets in the runaway regime may be far less efficient. \bdf{The key reason for this are complicated 3D gas flows that carry gas both in and out of the Hill sphere of the planet, resulting in a much smaller net planet accretion rate for Saturn-mass planets \citep[e.g.,][]{LambrechtsEtal19-No_CA-runaway}.} \cite{NayakshinEtal19} found in simplified (gas only, and no planet migration) models that suppressing gas accretion by a factor $F_{\rm acc} \sim 10$ was required to account for the predominance of sub-Jovian mass planets inferred to be present in ALMA discs \citep[][]{Dsharp7,LodatoEtal09}.

Our experiments here show that this conclusion also holds in models with planet migration and dust dynamics included. Planets with gas accretion suppressed by factor $F_{\rm acc} = 10$ allow planets to spend more time in the intermediate mass regime. Gaps opened by such planets in dusty discs are more moderate (cf. the green dashed curve in fig. \ref{fig:HD-Facc-Fmig-10}). However, at the full unquenched migration speeds these planets disappear into the inner disc too quickly, and the lifetime of the moderate gap phase, $t_{\rm gap}$, is extended only by a factor of $\sim 2$ (cf. the middle panel in fig. \ref{fig:Lodato19_sample_dt_histograms}).

We therefore also explored the models in which both gas accretion and planet migration is suppressed. This provides far more promising results. First of all, for slowly migrating planets the dust ring morphology switches from the ``inside ring" to the ``outside ring" \citep[see][and fig. \ref{fig:HD-Facc-Fmig-10}]{Meru19-Ring} morphology. The latter is generally invoked as explanation for the observed ALMA dust structures when one neglects planet migration \citep[e.g.,][]{DSHARP-6,Dsharp7}. The models with unsuppressed migration result instead in the ``inside ring" dust morphology, which, in addition to all the other problems already pointed out in this section, has so far not been invoked for any of the observed ALMA protoplanetary discs. Secondly, models with both growth and migration suppressed produce far longer lived moderate gap structures (the bottom panel in fig. \ref{fig:Lodato19_sample_dt_histograms}) that require a far smaller number of gas giants born per ALMA disc (the blue filled histogram in fig. \ref{fig:Lodato19_N_required}).

\subsection{Caveats}\label{sec:caveats}

\bdf{Here we made an assumption that the annular features seen in the dusty protoplanetary discs with ALMA are all due to embedded planets. This assumption is clearly an oversimplification since other physical processes were also shown to result in annular features, such as dead zones \citep{Ruge16_rings_by_dead_zones}, condensation fronts for important chemical species \citep{ZhangEtal15-cond-front},  dust self-induced pile-ups \citep{Gonzalez15_dust_pile_ups}, effects of vortices \citep[e.g.,][]{Barge17_dust_ring_vortices}, etc. We however believe that our conclusions are robust to at least a qualitative if not a quantitative level. We would expect a major change to our conclusions only if the majority of gaps we included in our sample are not due to planets. While this is in principle possible, presence of gas giant planets in protoplanetary discs has now been confirmed through independent means, such as localised velocity perturbations to gas flows in vicinity of planets \citep[e.g.,][]{Perez15_v_kinks,Perez18-Co-vkink,Pinte18-HD163296,Casassus19-HD100546,Teague19-HD163296,Pinte20-Dsharp-Vkinks}, H-$\alpha$ imaging \citep{HaffertEtal19}, and even detection of circum-planetary discs \citep{BenistyEtal21}. Nevertheless, more observational and theoretical/simulation work is needed to disentangle different mechanisms for formation of annular disc features. }

In \S \ref{sec:Fa10} we made the {\em assumption} that planet migration can be slowed down in the type I regime with regards to the well known formulae from \cite{Paardekooper11-typeI}. This allowed us to reduce the mismatch between the CA scenario predictions and ALMA observations. It is not clear whether this is reasonable. \cite{NelsonPap04} showed that MHD turbulence may result in far slower although chaotic type I migration of planets. However the values of the turbulent viscosity parameter they considered were larger than the upper limits on $\alpha_{\rm v}$ for the ALMA discs in which such measurements were possible.

There is some recent work showing that {\em luminous} planets can migrate slower or even outward due to ``thermal torques" \citep{BL15_thermal_torques,Masset17_thermal_torque}, but it appears that this effect is short lived \citep{Guilera21_thermal_torques} and is unlikely to reduce planet inward migration on time scales of $\gtrsim 1$ Myr; it also appears to affect planets only inside $10-20$ AU. \cite{Kimming20-wind-driven-migration} have demonstrated that planets migrating in the type II regime may be driven outward if the main mechanism of angular momentum transfer is due to MHD disc winds. However, we argued that we need type I migrating planets to account for the majority of ALMA annular features as type II migrating planets produce gaps that are too deep and wide. 

The last point \citep[on the potential importance of MHD driven disc winds, see, e.g.,][]{Tabone22-general} emphasises an important limitation of our study. Here we assumed that we know the protoplanetary disc physics and thus we can use ALMA observations to constrain CA scenario for planet formation. To be specific, we assumed here that the angular momentum transport is viscous and the disc is in a quasi steady state, which implies a minimum disc mass as explained in \S \ref{sec:ALMA_sample}. There can obviously be other complications such as non-steady accretion.
It is thus not impossible that the disc structure or its parameters, such as its mass, is significantly off from what we currently think it is. If that is true then perhaps the data can be made consistent with the classical runaway accretion scenario of CA and instead require a significant re-evaluation of our views of how the discs evolve.

\section{Conclusions}

In this paper we have shown that ALMA and other modern data of protoplanetary discs with embedded planet candidates present us with a unique opportunity to test planet formation models, supplementing constraints from the previous exoplanetary data of older discless, planetary systems. We constrained protoplanetary disc masses given their age, the presence of mm-sized grains, and the observed gas accretion rates onto the star. This then allows us to evaluate how  planets embedded in these discs should evolve for a specific planet formation model. As an example of this method, we examined the very successful Bern population synthesis implementation \citep{Bern20-1,Bern20-2} of the Core Accretion scenario.  Our results suggest that if the observed substructure is due to embedded planets, then both accretion and migration will probably need to be substantially slower than expected. We also emphasise that while we focused on the Bern model, our results would apply to any planet formation model that incorporates planet accretion and migration at rates similar to those in the Bern model. Finally, as noted in \S \ref{sec:caveats}, another possible interpretation of our results is that the data may require re-evaluation of our disc models rather than planet formation physics. In any event, we believe that our results indicate that the future of planet formation population synthesis is in applying it to data of both ALMA protoplanetary discs and the much older exoplanetary systems that have lost all signatures of their primordial discs.

\begin{table*}
\centering
\caption{\label{tab:1} The sample of ALMA discs and candidate planets investigated in the paper, and the inferred gap lifetime $t_{\rm gap}$ in units of Myr for the three scenarios. See \S \ref{sec:ALMA_sample} for detail. Except for the total disc mass, $M_{\rm disc}$, all the entries to the left of the vertical bar in the Table are those derived in the previous literature through observations or simulations to fit dusty disc morphology \citep[e.g.,][]{Dsharp7,LodatoEtal19}.  To the right of the vertical bar, the three columns list the results of our modelling, that is, the gap lifetimes for the Fiducial model, the one with gas accretion suppressed, and with both accretion and migration suppressed by factor of 10, respectively.}

\begin{tabular}{ccccccccc|ccc}
\hline 
\hline 
Object & $M_*$ & $L_*$ & Age & $M_{\rm disc}$ & $R_{\rm gap}$ & $\Delta_{\rm gap}$ & $C_{\rm gap}$ & $M_{\rm p}$ & $t_{\rm gap}$ & $t_{\rm gap}$ & $t_{\rm gap}$ \tabularnewline
& [$M_{\odot}$] & [$L_{\odot}$] & [Myr] & [$M_{\odot}$] & [AU] & [AU] & & [$\mj$] & Fiducial & $F_{\rm a} = 10, F_{\rm m} = 1$ &  $F_{\rm a} = F_{\rm m} = 10$ \tabularnewline
\hline 
DS TAU & 0.58 & 0.25 & 3.981 & 0.116 & 33 & 0.821 & 0.042 & 6.061 & 0.540 & 0.540 & 0.540 \tabularnewline
MWC480 & 1.9 & 17.4 & 6.310 & 0.382 & 73 & 0.454 & 0.013 & 2.300 & 0.166 & 0.194 & 0.241 \tabularnewline
GO TAU & 0.36 & 0.2 & 5.623 & 0.072 & 59 & 0.389 & 0.056 & 0.400 & 0.065 & 0.107 & 0.361 \tabularnewline
DL TAU & 0.98 & 0.65 & 1.000 & 0.059 & 67 & 0.207 & 0.157 & 0.163 & 0.032 & 0.225 & 1.035 \tabularnewline
DL TAU & 0.98 & 0.65 & 1.000 & 0.059 & 89 & 0.292 & 0.470 & 0.459 & 0.051 & 0.115 & 1.001 \tabularnewline
CI TAU & 0.89 & 0.8 & 1.585 & 0.104 & 14 & 0.636 & 0.455 & 0.750 & 0.239 & 0.239 & 0.239 \tabularnewline
CI TAU & 0.89 & 0.8 & 1.585 & 0.104 & 48 & 0.225 & 0.820 & 0.150 & 0.022 & 0.094 & 0.536 \tabularnewline
CI TAU & 0.89 & 0.8 & 1.585 & 0.104 & 119 & 0.186 & 0.600 & 0.400 & 0.028 & 0.077 & 0.705 \tabularnewline
HL TAU & 1.30 & 3.2 & 1.413 & 0.108 & 13 & 0.432 & 0.410 & 0.200 & 0.097 & 0.161 & 0.613 \tabularnewline
HL TAU & 1.30 & 3.2 & 1.413 & 0.108 & 32 & 0.152 & 0.670 & 0.270 & 0.041 & 0.097 & 0.636 \tabularnewline
HL TAU & 1.30 & 3.2 & 1.413 & 0.108 & 67 & 0.064 & 0.840 & 0.550 & 0.047 & 0.090 & 0.613 \tabularnewline
AS209 & 0.83 & 0.5 & 1.096 & 0.110 & 99 & 0.313 & 0.030 & 0.650 & 0.054 & 0.109 & 0.363 \tabularnewline
HD142666 & 1.58 & 9.1 & 12.589 & 0.316 & 16 & 0.219 & 0.730 & 0.300 & 0.051 & 0.123 & 0.230 \tabularnewline
HD143006 & 1.78 & 3.8 & 3.981 & 0.273 & 22 & 0.986 & 0.040 & 20.000 & 0.440 & 0.440 & 0.440 \tabularnewline
HD143006 & 1.78 & 3.8 & 3.981 & 0.273 & 51 & 0.251 & 0.530 & 0.330 & 0.027 & 0.068 & 0.257 \tabularnewline
HD163296 & 2.04 & 17.0 & 5.012 & 0.210 & 48 & 0.421 & 0.033 & 2.180 & 0.345 & 0.372 & 0.501 \tabularnewline
HD163296 & 2.04 & 17.0 & 5.012 & 0.210 & 86 & 0.188 & 0.150 & 1.000 & 0.045 & 0.090 & 0.379 \tabularnewline
SR4 & 0.68 & 1.2 & 0.80 & 0.136 & 11 & 0.573 & 0.230 & 2.160 & 0.235 & 0.235 & 0.235 \tabularnewline
HD169142 & 1.65 & 10.0 & 5.012 & 0.330 & 14 & 1.857 & 0.006 & 3.500 & 0.205 & 0.205 & 0.205 \tabularnewline
HD169142 & 1.65 & 10.0 & 5.012 & 0.330 & 40 & 0.750 & 0.012 & 0.700 & 0.061 & 0.097 & 0.142 \tabularnewline
PDS70 & 0.85 & 2.8 & 5.370 & 0.170 & 22 & 0.055 & 0.000 & 10.000 & 0.353 & 0.353 & 0.353 \tabularnewline

\hline 
\end{tabular}
\end{table*}

\section*{Acknowledgement}

Yinhao Wu is thanked for helping the authors to compare the migration rate of planets in the 1D code with the 2D code FARGO3D (see Appendix). The authors acknowledge funding from the UK Science and Technology Facilities Council under the grant No. ST/S000453/1. GR acknowledges support from an STFC Ernest Rutherford Fellowship (grant number ST/T003855/1). VE acknowledges support by the Ministry of Science and Higher Education of the Russian Federation under the grant 075-15-2020-780 (N13.1902.21.0039).

\section*{Data availability statement}

The data underlying this article will be shared on reasonable request to the corresponding author.


\bibliographystyle{mnras}
\bibliography{nayakshin}

\begin{thebibliography}{}
\makeatletter
\relax
\def\mn@urlcharsother{\let\do\@makeother \do\$\do\&\do\#\do\^\do\_\do\%\do\~}
\def\mn@doi{\begingroup\mn@urlcharsother \@ifnextchar [ {\mn@doi@}
  {\mn@doi@[]}}
\def\mn@doi@[#1]#2{\def\@tempa{#1}\ifx\@tempa\@empty \href
  {http://dx.doi.org/#2} {doi:#2}\else \href {http://dx.doi.org/#2} {#1}\fi
  \endgroup}
\def\mn@eprint#1#2{\mn@eprint@#1:#2::\@nil}
\def\mn@eprint@arXiv#1{\href {http://arxiv.org/abs/#1} {{\tt arXiv:#1}}}
\def\mn@eprint@dblp#1{\href {http://dblp.uni-trier.de/rec/bibtex/#1.xml}
  {dblp:#1}}
\def\mn@eprint@#1:#2:#3:#4\@nil{\def\@tempa {#1}\def\@tempb {#2}\def\@tempc
  {#3}\ifx \@tempc \@empty \let \@tempc \@tempb \let \@tempb \@tempa \fi \ifx
  \@tempb \@empty \def\@tempb {arXiv}\fi \@ifundefined
  {mn@eprint@\@tempb}{\@tempb:\@tempc}{\expandafter \expandafter \csname
  mn@eprint@\@tempb\endcsname \expandafter{\@tempc}}}

\bibitem[\protect\citeauthoryear{{ALMA Partnership} et~al.,}{{ALMA Partnership}
  et~al.}{2015}]{BroganEtal15}
{ALMA Partnership} et~al., 2015, \mn@doi [\apjl] {10.1088/2041-8205/808/1/L3},
  \href {http://adsabs.harvard.edu/abs/2015ApJ...808L...3A} {808, L3}

\bibitem[\protect\citeauthoryear{{Alexander}, {Clarke}  \&
  {Pringle}}{{Alexander} et~al.}{2006}]{AlexanderEtal06}
{Alexander} R.~D.,  {Clarke} C.~J.,   {Pringle} J.~E.,  2006, \mn@doi [\mnras]
  {10.1111/j.1365-2966.2006.10294.x}, \href
  {http://adsabs.harvard.edu/abs/2006MNRAS.369..229A} {369, 229}

\bibitem[\protect\citeauthoryear{{Alexander}, {Pascucci}, {Andrews}, {Armitage}
   \& {Cieza}}{{Alexander} et~al.}{2014}]{AlexanderREtal14a}
{Alexander} R.,  {Pascucci} I.,  {Andrews} S.,  {Armitage} P.,   {Cieza} L.,
  2014, \mn@doi [Protostars and Planets VI]
  {10.2458/azu_uapress_9780816531240-ch021}, \href
  {http://adsabs.harvard.edu/abs/2014prpl.conf..475A} {pp 475--496}

\bibitem[\protect\citeauthoryear{{Andrews}}{{Andrews}}{2020}]{Andrews20-Review}
{Andrews} S.~M.,  2020, \mn@doi [\araa] {10.1146/annurev-astro-031220-010302},
  \href {https://ui.adsabs.harvard.edu/abs/2020ARA&A..58..483A} {58, 483}

\bibitem[\protect\citeauthoryear{{Andrews} et~al.,}{{Andrews}
  et~al.}{2018}]{Dsharp1}
{Andrews} S.~M.,  et~al., 2018, \mn@doi [\apjl] {10.3847/2041-8213/aaf741},
  \href {http://adsabs.harvard.edu/abs/2018ApJ...869L..41A} {869, L41}

\bibitem[\protect\citeauthoryear{{Armitage} \& {Bonnell}}{{Armitage} \&
  {Bonnell}}{2002}]{armibonnell02}
{Armitage} P.~J.,  {Bonnell} I.~A.,  2002, \mn@doi [\mnras]
  {10.1046/j.1365-8711.2002.05213.x}, \href
  {http://adsabs.harvard.edu/abs/2002MNRAS.330L..11A} {330, L11}

\bibitem[\protect\citeauthoryear{{Barge}, {Ricci}, {Carilli}  \&
  {Previn-Ratnasingam}}{{Barge} et~al.}{2017}]{Barge17_dust_ring_vortices}
{Barge} P.,  {Ricci} L.,  {Carilli} C.~L.,   {Previn-Ratnasingam} R.,  2017,
  \mn@doi [\aap] {10.1051/0004-6361/201629918}, \href
  {https://ui.adsabs.harvard.edu/abs/2017A&A...605A.122B} {605, A122}

\bibitem[\protect\citeauthoryear{{Bate}, {Bonnell}  \& {Bromm}}{{Bate}
  et~al.}{2003}]{BateEtal03}
{Bate} M.~R.,  {Bonnell} I.~A.,   {Bromm} V.,  2003, \mn@doi [\mnras]
  {10.1046/j.1365-8711.2003.06210.x}, \href
  {http://ukads.nottingham.ac.uk/abs/2003MNRAS.339..577B} {339, 577}

\bibitem[\protect\citeauthoryear{{Benisty} et~al.,}{{Benisty}
  et~al.}{2021}]{BenistyEtal21}
{Benisty} M.,  et~al., 2021, \mn@doi [\apjl] {10.3847/2041-8213/ac0f83}, \href
  {https://ui.adsabs.harvard.edu/abs/2021ApJ...916L...2B} {916, L2}

\bibitem[\protect\citeauthoryear{{Ben{\'\i}tez-Llambay} \&
  {Masset}}{{Ben{\'\i}tez-Llambay} \& {Masset}}{2016}]{BL_Masset_16_FARGO3D}
{Ben{\'\i}tez-Llambay} P.,  {Masset} F.~S.,  2016, \mn@doi [\apjs]
  {10.3847/0067-0049/223/1/11}, \href
  {https://ui.adsabs.harvard.edu/abs/2016ApJS..223...11B} {223, 11}

\bibitem[\protect\citeauthoryear{{Ben{\'\i}tez-Llambay}, {Masset},
  {Koenigsberger}  \& {Szul{\'a}gyi}}{{Ben{\'\i}tez-Llambay}
  et~al.}{2015}]{BL15_thermal_torques}
{Ben{\'\i}tez-Llambay} P.,  {Masset} F.,  {Koenigsberger} G.,   {Szul{\'a}gyi}
  J.,  2015, \mn@doi [\nat] {10.1038/nature14277}, \href
  {https://ui.adsabs.harvard.edu/abs/2015Natur.520...63B} {520, 63}

\bibitem[\protect\citeauthoryear{{Bennett}, {Ranc}  \& {Fernandes}}{{Bennett}
  et~al.}{2021}]{Bennett21-No-runaway-desert}
{Bennett} D.~P.,  {Ranc} C.,   {Fernandes} R.~B.,  2021, \mn@doi [\aj]
  {10.3847/1538-3881/ac2a2b}, \href
  {https://ui.adsabs.harvard.edu/abs/2021AJ....162..243B} {162, 243}

\bibitem[\protect\citeauthoryear{{Birnstiel}, {Dullemond}  \&
  {Brauer}}{{Birnstiel} et~al.}{2009}]{Birnstiel09}
{Birnstiel} T.,  {Dullemond} C.~P.,   {Brauer} F.,  2009, \mn@doi [\aap]
  {10.1051/0004-6361/200912452}, \href
  {https://ui.adsabs.harvard.edu/abs/2009A%26A...503L...5B} {503, L5}

\bibitem[\protect\citeauthoryear{{Birnstiel}, {Klahr}  \&
  {Ercolano}}{{Birnstiel} et~al.}{2012}]{BirnstielEtal12}
{Birnstiel} T.,  {Klahr} H.,   {Ercolano} B.,  2012, \mn@doi [\aap]
  {10.1051/0004-6361/201118136}, \href
  {https://ui.adsabs.harvard.edu/abs/2012A%26A...539A.148B} {539, A148}

\bibitem[\protect\citeauthoryear{{Booth}, {Walsh}, {Ilee}, {Notsu}, {Qi},
  {Nomura}  \& {Akiyama}}{{Booth} et~al.}{2019}]{BoothEtal19-HD63296}
{Booth} A.~S.,  {Walsh} C.,  {Ilee} J.~D.,  {Notsu} S.,  {Qi} C.,  {Nomura} H.,
    {Akiyama} E.,  2019, \mn@doi [\apjl] {10.3847/2041-8213/ab3645}, \href
  {https://ui.adsabs.harvard.edu/abs/2019ApJ...882L..31B} {882, L31}

\bibitem[\protect\citeauthoryear{{Casassus} \& {P{\'e}rez}}{{Casassus} \&
  {P{\'e}rez}}{2019}]{Casassus19-HD100546}
{Casassus} S.,  {P{\'e}rez} S.,  2019, \mn@doi [\apjl]
  {10.3847/2041-8213/ab4425}, \href
  {https://ui.adsabs.harvard.edu/abs/2019ApJ...883L..41C} {883, L41}

\bibitem[\protect\citeauthoryear{{Coleman} \& {Nelson}}{{Coleman} \&
  {Nelson}}{2014}]{ColemanNelson14}
{Coleman} G. A.~L.,  {Nelson} R.~P.,  2014, \mn@doi [\mnras]
  {10.1093/mnras/stu1715}, \href
  {https://ui.adsabs.harvard.edu/abs/2014MNRAS.445..479C} {445, 479}

\bibitem[\protect\citeauthoryear{{Coleman} \& {Nelson}}{{Coleman} \&
  {Nelson}}{2016}]{ColemanNelson16}
{Coleman} G.~A.~L.,  {Nelson} R.~P.,  2016, \mn@doi [\mnras]
  {10.1093/mnras/stw1177}, \href
  {http://adsabs.harvard.edu/abs/2016MNRAS.460.2779C} {460, 2779}

\bibitem[\protect\citeauthoryear{{Crida} \& {Bitsch}}{{Crida} \&
  {Bitsch}}{2017}]{CridaBitsch-17-runaway-migration}
{Crida} A.,  {Bitsch} B.,  2017, \mn@doi [\icarus]
  {10.1016/j.icarus.2016.10.017}, \href
  {https://ui.adsabs.harvard.edu/abs/2017Icar..285..145C} {285, 145}

\bibitem[\protect\citeauthoryear{{Crida} \& {Morbidelli}}{{Crida} \&
  {Morbidelli}}{2007}]{CridaMorbidelli07}
{Crida} A.,  {Morbidelli} A.,  2007, \mn@doi [\mnras]
  {10.1111/j.1365-2966.2007.11704.x}, \href
  {http://adsabs.harvard.edu/abs/2007MNRAS.377.1324C} {377, 1324}

\bibitem[\protect\citeauthoryear{{Crida}, {Morbidelli}  \& {Masset}}{{Crida}
  et~al.}{2006}]{CridaEtal06}
{Crida} A.,  {Morbidelli} A.,   {Masset} F.,  2006, \mn@doi [\icarus]
  {10.1016/j.icarus.2005.10.007}, \href
  {http://adsabs.harvard.edu/abs/2006Icar..181..587C} {181, 587}

\bibitem[\protect\citeauthoryear{{D'Alessio}, {Calvet}  \&
  {Hartmann}}{{D'Alessio} et~al.}{2001}]{DAlessioEtal01}
{D'Alessio} P.,  {Calvet} N.,   {Hartmann} L.,  2001, \mn@doi [\apj]
  {10.1086/320655}, \href {http://adsabs.harvard.edu/abs/2001ApJ...553..321D}
  {553, 321}

\bibitem[\protect\citeauthoryear{{Dipierro} \& {Laibe}}{{Dipierro} \&
  {Laibe}}{2017}]{DipierroLaibe17}
{Dipierro} G.,  {Laibe} G.,  2017, \mn@doi [\mnras] {10.1093/mnras/stx977},
  \href {http://adsabs.harvard.edu/abs/2017MNRAS.469.1932D} {469, 1932}

\bibitem[\protect\citeauthoryear{{Dipierro}, {Laibe}, {Price}  \&
  {Lodato}}{{Dipierro} et~al.}{2016}]{DipierroEtal16a}
{Dipierro} G.,  {Laibe} G.,  {Price} D.~J.,   {Lodato} G.,  2016, \mn@doi
  [\mnras] {10.1093/mnrasl/slw032}, \href
  {http://adsabs.harvard.edu/abs/2016MNRAS.tmpL..16D} {}

\bibitem[\protect\citeauthoryear{{Dullemond} et~al.,}{{Dullemond}
  et~al.}{2018}]{DSHARP-6}
{Dullemond} C.~P.,  et~al., 2018, \mn@doi [\apjl] {10.3847/2041-8213/aaf742},
  \href {http://adsabs.harvard.edu/abs/2018ApJ...869L..46D} {869, L46}

\bibitem[\protect\citeauthoryear{{Dullemond}, {Isella}, {Andrews}, {Skobleva}
  \& {Dzyurkevich}}{{Dullemond} et~al.}{2020}]{DullemondEtal20-HD163296}
{Dullemond} C.~P.,  {Isella} A.,  {Andrews} S.~M.,  {Skobleva} I.,
  {Dzyurkevich} N.,  2020, \mn@doi [\aap] {10.1051/0004-6361/201936438}, \href
  {https://ui.adsabs.harvard.edu/abs/2020A&A...633A.137D} {633, A137}

\bibitem[\protect\citeauthoryear{{Emsenhuber}, {Mordasini}, {Burn}, {Alibert},
  {Benz}  \& {Asphaug}}{{Emsenhuber} et~al.}{2021a}]{Bern20-1}
{Emsenhuber} A.,  {Mordasini} C.,  {Burn} R.,  {Alibert} Y.,  {Benz} W.,
  {Asphaug} E.,  2021a, \mn@doi [\aap] {10.1051/0004-6361/202038553}, \href
  {https://ui.adsabs.harvard.edu/abs/2021A&A...656A..69E} {656, A69}

\bibitem[\protect\citeauthoryear{{Emsenhuber}, {Mordasini}, {Burn}, {Alibert},
  {Benz}  \& {Asphaug}}{{Emsenhuber} et~al.}{2021b}]{Bern20-2}
{Emsenhuber} A.,  {Mordasini} C.,  {Burn} R.,  {Alibert} Y.,  {Benz} W.,
  {Asphaug} E.,  2021b, \mn@doi [\aap] {10.1051/0004-6361/202038863}, \href
  {https://ui.adsabs.harvard.edu/abs/2021A&A...656A..70E} {656, A70}

\bibitem[\protect\citeauthoryear{{Flaherty} et~al.,}{{Flaherty}
  et~al.}{2017}]{Flaherty17-HD163296-turb}
{Flaherty} K.~M.,  et~al., 2017, \mn@doi [\apj] {10.3847/1538-4357/aa79f9},
  \href {https://ui.adsabs.harvard.edu/abs/2017ApJ...843..150F} {843, 150}

\bibitem[\protect\citeauthoryear{{Fung} \& {Chiang}}{{Fung} \&
  {Chiang}}{2016}]{FungChiang16}
{Fung} J.,  {Chiang} E.,  2016, \mn@doi [\apj] {10.3847/0004-637X/832/2/105},
  \href {http://esoads.eso.org/abs/2016ApJ...832..105F} {832, 105}

\bibitem[\protect\citeauthoryear{{Gonzalez}, {Laibe}, {Maddison}, {Pinte}  \&
  {M{\'e}nard}}{{Gonzalez} et~al.}{2015}]{Gonzalez15_dust_pile_ups}
{Gonzalez} J.~F.,  {Laibe} G.,  {Maddison} S.~T.,  {Pinte} C.,   {M{\'e}nard}
  F.,  2015, \mn@doi [\mnras] {10.1093/mnrasl/slv120}, \href
  {https://ui.adsabs.harvard.edu/abs/2015MNRAS.454L..36G} {454, L36}

\bibitem[\protect\citeauthoryear{{Guilera}, {Miller Bertolami}, {Masset},
  {Cuadra}, {Venturini}  \& {Ronco}}{{Guilera}
  et~al.}{2021}]{Guilera21_thermal_torques}
{Guilera} O.~M.,  {Miller Bertolami} M.~M.,  {Masset} F.,  {Cuadra} J.,
  {Venturini} J.,   {Ronco} M.~P.,  2021, \mn@doi [\mnras]
  {10.1093/mnras/stab2371}, \href
  {https://ui.adsabs.harvard.edu/abs/2021MNRAS.507.3638G} {507, 3638}

\bibitem[\protect\citeauthoryear{{Haffert}, {Bohn}, {de Boer}, {Snellen},
  {Brinchmann}, {Girard}, {Keller}  \& {Bacon}}{{Haffert}
  et~al.}{2019}]{HaffertEtal19}
{Haffert} S.~Y.,  {Bohn} A.~J.,  {de Boer} J.,  {Snellen} I.~A.~G.,
  {Brinchmann} J.,  {Girard} J.~H.,  {Keller} C.~U.,   {Bacon} R.,  2019,
  \mn@doi [Nature Astronomy] {10.1038/s41550-019-0780-5}, \href
  {https://ui.adsabs.harvard.edu/abs/2019NatAs...3..749H} {3, 749}

\bibitem[\protect\citeauthoryear{{Hall}, {Dong}, {Rice}, {Harries}, {Najita},
  {Alexander}  \& {Brittain}}{{Hall} et~al.}{2019}]{HallEtal18}
{Hall} C.,  {Dong} R.,  {Rice} K.,  {Harries} T.~J.,  {Najita} J.,  {Alexander}
  R.,   {Brittain} S.,  2019, \mn@doi [\apj] {10.3847/1538-4357/aafac2}, \href
  {https://ui.adsabs.harvard.edu/abs/2019ApJ...871..228H} {871, 228}

\bibitem[\protect\citeauthoryear{{Huang} et~al.,}{{Huang}
  et~al.}{2018}]{Dsharp2}
{Huang} J.,  et~al., 2018, \mn@doi [\apjl] {10.3847/2041-8213/aaf740}, \href
  {http://adsabs.harvard.edu/abs/2018ApJ...869L..42H} {869, L42}

\bibitem[\protect\citeauthoryear{{Ida} \& {Lin}}{{Ida} \&
  {Lin}}{2004a}]{IdaLin04a}
{Ida} S.,  {Lin} D.~N.~C.,  2004a, \mn@doi [\apj] {10.1086/381724}, \href
  {http://adsabs.harvard.edu/abs/2004ApJ...604..388I} {604, 388}

\bibitem[\protect\citeauthoryear{{Ida} \& {Lin}}{{Ida} \&
  {Lin}}{2004b}]{IdaLin04b}
{Ida} S.,  {Lin} D.~N.~C.,  2004b, \mn@doi [\apj] {10.1086/424830}, \href
  {http://ukads.nottingham.ac.uk/abs/2004ApJ...616..567I} {616, 567}

\bibitem[\protect\citeauthoryear{{Ikoma}, {Nakazawa}  \& {Emori}}{{Ikoma}
  et~al.}{2000}]{IkomaEtal00}
{Ikoma} M.,  {Nakazawa} K.,   {Emori} H.,  2000, \mn@doi [\apj]
  {10.1086/309050}, \href
  {http://ukads.nottingham.ac.uk/abs/2000ApJ...537.1013I} {537, 1013}

\bibitem[\protect\citeauthoryear{{Jones}, {Pringle}  \& {Alexander}}{{Jones}
  et~al.}{2012}]{JPA12}
{Jones} M.~G.,  {Pringle} J.~E.,   {Alexander} R.~D.,  2012, \mn@doi [\mnras]
  {10.1111/j.1365-2966.2011.19730.x}, \href
  {http://adsabs.harvard.edu/abs/2012MNRAS.419..925J} {419, 925}

\bibitem[\protect\citeauthoryear{{Jung} et~al.,}{{Jung}
  et~al.}{2019}]{Jung21-microlensing-PMF}
{Jung} Y.~K.,  et~al., 2019, \mn@doi [\aj] {10.3847/1538-3881/aaf87f}, \href
  {https://ui.adsabs.harvard.edu/abs/2019AJ....157...72J} {157, 72}

\bibitem[\protect\citeauthoryear{{Keppler} et~al.,}{{Keppler}
  et~al.}{2019}]{KepplerEtal19}
{Keppler} M.,  et~al., 2019, \mn@doi [\aap] {10.1051/0004-6361/201935034},
  \href {https://ui.adsabs.harvard.edu/abs/2019A&A...625A.118K} {625, A118}

\bibitem[\protect\citeauthoryear{{Kimmig}, {Dullemond}  \& {Kley}}{{Kimmig}
  et~al.}{2020}]{Kimming20-wind-driven-migration}
{Kimmig} C.~N.,  {Dullemond} C.~P.,   {Kley} W.,  2020, \mn@doi [\aap]
  {10.1051/0004-6361/201936412}, \href
  {https://ui.adsabs.harvard.edu/abs/2020A&A...633A...4K} {633, A4}

\bibitem[\protect\citeauthoryear{{Lambrechts} \& {Lega}}{{Lambrechts} \&
  {Lega}}{2017}]{LambrechtsLega17}
{Lambrechts} M.,  {Lega} E.,  2017, \mn@doi [\aap]
  {10.1051/0004-6361/201731014}, \href
  {http://adsabs.harvard.edu/abs/2017A%26A...606A.146L} {606, A146}

\bibitem[\protect\citeauthoryear{{Lambrechts}, {Lega}, {Nelson}, {Crida}  \&
  {Morbidelli}}{{Lambrechts} et~al.}{2019}]{LambrechtsEtal19-No_CA-runaway}
{Lambrechts} M.,  {Lega} E.,  {Nelson} R.~P.,  {Crida} A.,   {Morbidelli} A.,
  2019, \mn@doi [\aap] {10.1051/0004-6361/201834413}, \href
  {https://ui.adsabs.harvard.edu/abs/2019A&A...630A..82L} {630, A82}

\bibitem[\protect\citeauthoryear{{Lin} \& {Papaloizou}}{{Lin} \&
  {Papaloizou}}{1986}]{LinPap86}
{Lin} D.~N.~C.,  {Papaloizou} J.,  1986, \mn@doi [\apj] {10.1086/164653}, \href
  {http://adsabs.harvard.edu/abs/1986ApJ...309..846L} {309, 846}

\bibitem[\protect\citeauthoryear{{Liu} et~al.,}{{Liu} et~al.}{2019}]{LiuEtal19}
{Liu} Y.,  et~al., 2019, \mn@doi [\aap] {10.1051/0004-6361/201834157}, \href
  {https://ui.adsabs.harvard.edu/abs/2019A%26A...622A..75L} {622, A75}

\bibitem[\protect\citeauthoryear{{Lodato} \& {Clarke}}{{Lodato} \&
  {Clarke}}{2004}]{LodatoClarke04}
{Lodato} G.,  {Clarke} C.~J.,  2004, \mn@doi [\mnras]
  {10.1111/j.1365-2966.2004.08112.x}, \href
  {http://adsabs.harvard.edu/abs/2004MNRAS.353..841L} {353, 841}

\bibitem[\protect\citeauthoryear{{Lodato}, {Nayakshin}, {King}  \&
  {Pringle}}{{Lodato} et~al.}{2009}]{LodatoEtal09}
{Lodato} G.,  {Nayakshin} S.,  {King} A.~R.,   {Pringle} J.~E.,  2009, \mn@doi
  [\mnras] {10.1111/j.1365-2966.2009.15179.x}, \href
  {http://adsabs.harvard.edu/abs/2009MNRAS.398.1392L} {398, 1392}

\bibitem[\protect\citeauthoryear{{Lodato} et~al.,}{{Lodato}
  et~al.}{2019}]{LodatoEtal19}
{Lodato} G.,  et~al., 2019, \mn@doi [\mnras] {10.1093/mnras/stz913}, \href
  {http://adsabs.harvard.edu/abs/2019MNRAS.486..453L} {486, 453}

\bibitem[\protect\citeauthoryear{{Long} et~al.,}{{Long}
  et~al.}{2018}]{LongEtal18}
{Long} F.,  et~al., 2018, \mn@doi [\apj] {10.3847/1538-4357/aae8e1}, \href
  {https://ui.adsabs.harvard.edu/abs/2018ApJ...869...17L} {869, 17}

\bibitem[\protect\citeauthoryear{{Manara} et~al.,}{{Manara}
  et~al.}{2017}]{ManaraT-etal17}
{Manara} C.~F.,  et~al., 2017, \mn@doi [\aap] {10.1051/0004-6361/201630147},
  \href {http://adsabs.harvard.edu/abs/2017A%26A...604A.127M} {604, A127}

\bibitem[\protect\citeauthoryear{{Masset}}{{Masset}}{2017}]{Masset17_thermal_torque}
{Masset} F.~S.,  2017, \mn@doi [\mnras] {10.1093/mnras/stx2271}, \href
  {https://ui.adsabs.harvard.edu/abs/2017MNRAS.472.4204M} {472, 4204}

\bibitem[\protect\citeauthoryear{{Mayor} et~al.,}{{Mayor}
  et~al.}{2011}]{MayorEtal11}
{Mayor} M.,  et~al., 2011, ArXiv e-prints (astro-ph 1109.2497), \href
  {http://adsabs.harvard.edu/abs/2011arXiv1109.2497M} {}

\bibitem[\protect\citeauthoryear{{Meru}, {Rosotti}, {Booth}, {Nazari}  \&
  {Clarke}}{{Meru} et~al.}{2019}]{Meru19-Ring}
{Meru} F.,  {Rosotti} G.~P.,  {Booth} R.~A.,  {Nazari} P.,   {Clarke} C.~J.,
  2019, \mn@doi [\mnras] {10.1093/mnras/sty2847}, \href
  {https://ui.adsabs.harvard.edu/abs/2019MNRAS.482.3678M} {482, 3678}

\bibitem[\protect\citeauthoryear{{Mizuno}}{{Mizuno}}{1980}]{Mizuno80}
{Mizuno} H.,  1980, \mn@doi [Progress of Theoretical Physics]
  {10.1143/PTP.64.544}, \href
  {http://ukads.nottingham.ac.uk/abs/1980PThPh..64..544M} {64, 544}

\bibitem[\protect\citeauthoryear{{Mordasini}, {Alibert}  \& {Benz}}{{Mordasini}
  et~al.}{2009a}]{MordasiniEtal09a}
{Mordasini} C.,  {Alibert} Y.,   {Benz} W.,  2009a, \mn@doi [\aap]
  {10.1051/0004-6361/200810301}, \href
  {http://adsabs.harvard.edu/abs/2009A%26A...501.1139M} {501, 1139}

\bibitem[\protect\citeauthoryear{{Mordasini}, {Alibert}, {Benz}  \&
  {Naef}}{{Mordasini} et~al.}{2009b}]{MordasiniEtal09b}
{Mordasini} C.,  {Alibert} Y.,  {Benz} W.,   {Naef} D.,  2009b, \mn@doi [\aap]
  {10.1051/0004-6361/200810697}, \href
  {http://adsabs.harvard.edu/abs/2009A%26A...501.1161M} {501, 1161}

\bibitem[\protect\citeauthoryear{{Mordasini}, {Alibert}, {Benz}, {Klahr}  \&
  {Henning}}{{Mordasini} et~al.}{2012}]{MordasiniEtal12}
{Mordasini} C.,  {Alibert} Y.,  {Benz} W.,  {Klahr} H.,   {Henning} T.,  2012,
  \mn@doi [\aap] {10.1051/0004-6361/201117350}, \href
  {http://adsabs.harvard.edu/abs/2012A%26A...541A..97M} {541, A97}

\bibitem[\protect\citeauthoryear{{Muro-Arena} et~al.,}{{Muro-Arena}
  et~al.}{2018}]{MuroA-Etal18-HD163296}
{Muro-Arena} G.~A.,  et~al., 2018, \mn@doi [\aap]
  {10.1051/0004-6361/201732299}, \href
  {https://ui.adsabs.harvard.edu/abs/2018A&A...614A..24M} {614, A24}

\bibitem[\protect\citeauthoryear{{Nayakshin} \& {Fletcher}}{{Nayakshin} \&
  {Fletcher}}{2015}]{NayakshinFletcher15}
{Nayakshin} S.,  {Fletcher} M.,  2015, \mn@doi [\mnras]
  {10.1093/mnras/stv1354}, \href
  {http://adsabs.harvard.edu/abs/2015MNRAS.452.1654N} {452, 1654}

\bibitem[\protect\citeauthoryear{{Nayakshin}, {Dipierro}  \&
  {Szul{\'a}gyi}}{{Nayakshin} et~al.}{2019}]{NayakshinEtal19}
{Nayakshin} S.,  {Dipierro} G.,   {Szul{\'a}gyi} J.,  2019, \mn@doi [\mnras]
  {10.1093/mnrasl/slz087}, \href
  {https://ui.adsabs.harvard.edu/abs/2019MNRAS.488L..12N} {488, L12}

\bibitem[\protect\citeauthoryear{{Ndugu}, {Bitsch}  \& {Jurua}}{{Ndugu}
  et~al.}{2019}]{NduguEtal19}
{Ndugu} N.,  {Bitsch} B.,   {Jurua} E.,  2019, \mn@doi [\mnras]
  {10.1093/mnras/stz1862}, \href
  {https://ui.adsabs.harvard.edu/abs/2019MNRAS.488.3625N} {488, 3625}

\bibitem[\protect\citeauthoryear{{Nelson} \& {Papaloizou}}{{Nelson} \&
  {Papaloizou}}{2004}]{NelsonPap04}
{Nelson} R.~P.,  {Papaloizou} J. C.~B.,  2004, \mn@doi [\mnras]
  {10.1111/j.1365-2966.2004.07406.x}, \href
  {https://ui.adsabs.harvard.edu/abs/2004MNRAS.350..849N} {350, 849}

\bibitem[\protect\citeauthoryear{{Ormel}, {Shi}  \& {Kuiper}}{{Ormel}
  et~al.}{2015}]{OrmelEtal15}
{Ormel} C.~W.,  {Shi} J.-M.,   {Kuiper} R.,  2015, \mn@doi [\mnras]
  {10.1093/mnras/stu2704}, \href
  {http://adsabs.harvard.edu/abs/2015MNRAS.447.3512O} {447, 3512}

\bibitem[\protect\citeauthoryear{{Paardekooper}, {Baruteau}, {Crida}  \&
  {Kley}}{{Paardekooper} et~al.}{2010}]{PaardekooperEtal10a}
{Paardekooper} S.-J.,  {Baruteau} C.,  {Crida} A.,   {Kley} W.,  2010, \mn@doi
  [\mnras] {10.1111/j.1365-2966.2009.15782.x}, \href
  {http://adsabs.harvard.edu/abs/2010MNRAS.401.1950P} {401, 1950}

\bibitem[\protect\citeauthoryear{{Paardekooper}, {Baruteau}  \&
  {Kley}}{{Paardekooper} et~al.}{2011}]{Paardekooper11-typeI}
{Paardekooper} S.~J.,  {Baruteau} C.,   {Kley} W.,  2011, \mn@doi [\mnras]
  {10.1111/j.1365-2966.2010.17442.x}, \href
  {https://ui.adsabs.harvard.edu/abs/2011MNRAS.410..293P} {410, 293}

\bibitem[\protect\citeauthoryear{{Perez}, {Dunhill}, {Casassus}, {Roman},
  {Szul{\'a}gyi}, {Flores}, {Marino}  \& {Montesinos}}{{Perez}
  et~al.}{2015}]{Perez15_v_kinks}
{Perez} S.,  {Dunhill} A.,  {Casassus} S.,  {Roman} P.,  {Szul{\'a}gyi} J.,
  {Flores} C.,  {Marino} S.,   {Montesinos} M.,  2015, \mn@doi [\apjl]
  {10.1088/2041-8205/811/1/L5}, \href
  {https://ui.adsabs.harvard.edu/abs/2015ApJ...811L...5P} {811, L5}

\bibitem[\protect\citeauthoryear{{P{\'e}rez}, {Casassus}  \&
  {Ben{\'\i}tez-Llambay}}{{P{\'e}rez} et~al.}{2018}]{Perez18-Co-vkink}
{P{\'e}rez} S.,  {Casassus} S.,   {Ben{\'\i}tez-Llambay} P.,  2018, \mn@doi
  [\mnras] {10.1093/mnrasl/sly109}, \href
  {https://ui.adsabs.harvard.edu/abs/2018MNRAS.480L..12P} {480, L12}

\bibitem[\protect\citeauthoryear{{Pinilla}, {Birnstiel}, {Ricci}, {Dullemond},
  {Uribe}, {Testi}  \& {Natta}}{{Pinilla} et~al.}{2012}]{PinillaEtal12}
{Pinilla} P.,  {Birnstiel} T.,  {Ricci} L.,  {Dullemond} C.~P.,  {Uribe} A.~L.,
   {Testi} L.,   {Natta} A.,  2012, \mn@doi [\aap]
  {10.1051/0004-6361/201118204}, \href
  {http://adsabs.harvard.edu/abs/2012A%26A...538A.114P} {538, A114}

\bibitem[\protect\citeauthoryear{{Pinte} et~al.,}{{Pinte}
  et~al.}{2018}]{Pinte18-HD163296}
{Pinte} C.,  et~al., 2018, \mn@doi [\apjl] {10.3847/2041-8213/aac6dc}, \href
  {https://ui.adsabs.harvard.edu/abs/2018ApJ...860L..13P} {860, L13}

\bibitem[\protect\citeauthoryear{{Pinte} et~al.,}{{Pinte}
  et~al.}{2020}]{Pinte20-Dsharp-Vkinks}
{Pinte} C.,  et~al., 2020, \mn@doi [\apjl] {10.3847/2041-8213/ab6dda}, \href
  {https://ui.adsabs.harvard.edu/abs/2020ApJ...890L...9P} {890, L9}

\bibitem[\protect\citeauthoryear{{Piso} \& {Youdin}}{{Piso} \&
  {Youdin}}{2014}]{PisoYoudin14}
{Piso} A.-M.~A.,  {Youdin} A.~N.,  2014, \mn@doi [\apj]
  {10.1088/0004-637X/786/1/21}, \href
  {http://adsabs.harvard.edu/abs/2014ApJ...786...21P} {786, 21}

\bibitem[\protect\citeauthoryear{{Piso}, {Youdin}  \& {Murray-Clay}}{{Piso}
  et~al.}{2015}]{Piso15}
{Piso} A.-M.~A.,  {Youdin} A.~N.,   {Murray-Clay} R.~A.,  2015, \mn@doi [\apj]
  {10.1088/0004-637X/800/2/82}, \href
  {http://esoads.eso.org/abs/2015ApJ...800...82P} {800, 82}

\bibitem[\protect\citeauthoryear{{Pollack}, {Hubickyj}, {Bodenheimer},
  {Lissauer}, {Podolak}  \& {Greenzweig}}{{Pollack}
  et~al.}{1996}]{PollackEtal96}
{Pollack} J.~B.,  {Hubickyj} O.,  {Bodenheimer} P.,  {Lissauer} J.~J.,
  {Podolak} M.,   {Greenzweig} Y.,  1996, \mn@doi [Icarus]
  {10.1006/icar.1996.0190}, \href
  {http://adsabs.harvard.edu/abs/1996Icar..124...62P} {124, 62}

\bibitem[\protect\citeauthoryear{{Powell}, {Murray-Clay}  \&
  {Schlichting}}{{Powell} et~al.}{2017}]{Powell17-dust-lines}
{Powell} D.,  {Murray-Clay} R.,   {Schlichting} H.~E.,  2017, \mn@doi [\apj]
  {10.3847/1538-4357/aa6d7c}, \href
  {https://ui.adsabs.harvard.edu/abs/2017ApJ...840...93P} {840, 93}

\bibitem[\protect\citeauthoryear{{Powell}, {Murray-Clay}, {P{\'e}rez},
  {Schlichting}  \& {Rosenthal}}{{Powell} et~al.}{2019}]{Powell19-DustLanes}
{Powell} D.,  {Murray-Clay} R.,  {P{\'e}rez} L.~M.,  {Schlichting} H.~E.,
  {Rosenthal} M.,  2019, \mn@doi [\apj] {10.3847/1538-4357/ab20ce}, \href
  {https://ui.adsabs.harvard.edu/abs/2019ApJ...878..116P} {878, 116}

\bibitem[\protect\citeauthoryear{{Qi}, {D'Alessio}, {{\"O}berg}, {Wilner},
  {Hughes}, {Andrews}  \& {Ayala}}{{Qi} et~al.}{2011}]{QiEtal11}
{Qi} C.,  {D'Alessio} P.,  {{\"O}berg} K.~I.,  {Wilner} D.~J.,  {Hughes} A.~M.,
   {Andrews} S.~M.,   {Ayala} S.,  2011, \mn@doi [\apj]
  {10.1088/0004-637X/740/2/84}, \href
  {https://ui.adsabs.harvard.edu/abs/2011ApJ...740...84Q} {740, 84}

\bibitem[\protect\citeauthoryear{{Rab}, {Kamp}, {Dominik}, {Ginski},
  {Muro-Arena}, {Thi}, {Waters}  \& {Woitke}}{{Rab}
  et~al.}{2020}]{RabEtal20-Temperature-in-gaps}
{Rab} C.,  {Kamp} I.,  {Dominik} C.,  {Ginski} C.,  {Muro-Arena} G.~A.,  {Thi}
  W.~F.,  {Waters} L.~B.~F.~M.,   {Woitke} P.,  2020, \mn@doi [\aap]
  {10.1051/0004-6361/202038712}, \href
  {https://ui.adsabs.harvard.edu/abs/2020A&A...642A.165R} {642, A165}

\bibitem[\protect\citeauthoryear{{Rice}, {Armitage}, {Wood}  \&
  {Lodato}}{{Rice} et~al.}{2006}]{RiceEtal06}
{Rice} W.~K.~M.,  {Armitage} P.~J.,  {Wood} K.,   {Lodato} G.,  2006, \mn@doi
  [\mnras] {10.1111/j.1365-2966.2006.11113.x}, \href
  {http://adsabs.harvard.edu/abs/2006MNRAS.373.1619R} {373, 1619}

\bibitem[\protect\citeauthoryear{{Rosotti}, {Juhasz}, {Booth}  \&
  {Clarke}}{{Rosotti} et~al.}{2016}]{RosottiEtal16}
{Rosotti} G.~P.,  {Juhasz} A.,  {Booth} R.~A.,   {Clarke} C.~J.,  2016, \mn@doi
  [\mnras] {10.1093/mnras/stw691}, \href
  {http://adsabs.harvard.edu/abs/2016MNRAS.459.2790R} {459, 2790}

\bibitem[\protect\citeauthoryear{{Rosotti}, {Tazzari}, {Booth}, {Testi},
  {Lodato}  \& {Clarke}}{{Rosotti} et~al.}{2019}]{Rosotti19-Opacity-Cliff}
{Rosotti} G.~P.,  {Tazzari} M.,  {Booth} R.~A.,  {Testi} L.,  {Lodato} G.,
  {Clarke} C.,  2019, \mn@doi [\mnras] {10.1093/mnras/stz1190}, \href
  {https://ui.adsabs.harvard.edu/abs/2019MNRAS.486.4829R} {486, 4829}

\bibitem[\protect\citeauthoryear{{Ruge}, {Flock}, {Wolf}, {Dzyurkevich},
  {Fromang}, {Henning}, {Klahr}  \& {Meheut}}{{Ruge}
  et~al.}{2016}]{Ruge16_rings_by_dead_zones}
{Ruge} J.~P.,  {Flock} M.,  {Wolf} S.,  {Dzyurkevich} N.,  {Fromang} S.,
  {Henning} T.,  {Klahr} H.,   {Meheut} H.,  2016, \mn@doi [\aap]
  {10.1051/0004-6361/201526616}, \href
  {https://ui.adsabs.harvard.edu/abs/2016A&A...590A..17R} {590, A17}

\bibitem[\protect\citeauthoryear{{Safronov}}{{Safronov}}{1969}]{Safronov69}
{Safronov} V.~S.,  1969, {Evoliutsiia doplanetnogo oblaka.}

\bibitem[\protect\citeauthoryear{{Shakura} \& {Sunyaev}}{{Shakura} \&
  {Sunyaev}}{1973}]{Shakura73}
{Shakura} N.~I.,  {Sunyaev} R.~A.,  1973, \aap, \href
  {http://cdsads.u-strasbg.fr/cgi-bin/nph-bib_query?bibcode=1973A%26A....24..337S&db_key=AST}
  {24, 337}

\bibitem[\protect\citeauthoryear{{Stevenson}}{{Stevenson}}{1982}]{Stevenson82}
{Stevenson} D.~J.,  1982, \mn@doi [P\&SS] {10.1016/0032-0633(82)90108-8}, \href
  {http://ukads.nottingham.ac.uk/abs/1982P%26SS...30..755S} {30, 755}

\bibitem[\protect\citeauthoryear{{Suzuki} et~al.,}{{Suzuki}
  et~al.}{2016}]{SuzukiEtal16}
{Suzuki} D.,  et~al., 2016, \mn@doi [\apj] {10.3847/1538-4357/833/2/145}, \href
  {http://adsabs.harvard.edu/abs/2016ApJ...833..145S} {833, 145}

\bibitem[\protect\citeauthoryear{{Suzuki} et~al.,}{{Suzuki}
  et~al.}{2018}]{SuzukiEtal18}
{Suzuki} D.,  et~al., 2018, \mn@doi [\apjl] {10.3847/2041-8213/aaf577}, \href
  {http://adsabs.harvard.edu/abs/2018ApJ...869L..34S} {869, L34}

\bibitem[\protect\citeauthoryear{{Syer} \& {Clarke}}{{Syer} \&
  {Clarke}}{1995}]{SyerClarke95}
{Syer} D.,  {Clarke} C.~J.,  1995, \mnras, \href
  {http://adsabs.harvard.edu/abs/1995MNRAS.277..758S} {277, 758}

\bibitem[\protect\citeauthoryear{{Szul{\'a}gyi}, {Morbidelli}, {Crida}  \&
  {Masset}}{{Szul{\'a}gyi} et~al.}{2014}]{Szulagyi14}
{Szul{\'a}gyi} J.,  {Morbidelli} A.,  {Crida} A.,   {Masset} F.,  2014, \mn@doi
  [\apj] {10.1088/0004-637X/782/2/65}, \href
  {http://esoads.eso.org/abs/2014ApJ...782...65S} {782, 65}

\bibitem[\protect\citeauthoryear{{Tabone}, {Rosotti}, {Cridland}, {Armitage}
  \& {Lodato}}{{Tabone} et~al.}{2021}]{Tabone22-general}
{Tabone} B.,  {Rosotti} G.~P.,  {Cridland} A.~J.,  {Armitage} P.~J.,   {Lodato}
  G.,  2021, \mn@doi [\mnras] {10.1093/mnras/stab3442}, \href
  {https://ui.adsabs.harvard.edu/abs/2021MNRAS.tmp.3115T} {}

\bibitem[\protect\citeauthoryear{{Teague}, {Bae}  \& {Bergin}}{{Teague}
  et~al.}{2019}]{Teague19-HD163296}
{Teague} R.,  {Bae} J.,   {Bergin} E.~A.,  2019, \mn@doi [\nat]
  {10.1038/s41586-019-1642-0}, \href
  {https://ui.adsabs.harvard.edu/abs/2019Natur.574..378T} {574, 378}

\bibitem[\protect\citeauthoryear{{Williams} \& {Cieza}}{{Williams} \&
  {Cieza}}{2011}]{WilliamsCiezaR11}
{Williams} J.~P.,  {Cieza} L.~A.,  2011, \mn@doi [\araa]
  {10.1146/annurev-astro-081710-102548}, \href
  {https://ui.adsabs.harvard.edu/abs/2011ARA&A..49...67W} {49, 67}

\bibitem[\protect\citeauthoryear{{Winn} \& {Fabrycky}}{{Winn} \&
  {Fabrycky}}{2015}]{WF14}
{Winn} J.~N.,  {Fabrycky} D.~C.,  2015, \mn@doi [\araa]
  {10.1146/annurev-astro-082214-122246}, \href
  {http://adsabs.harvard.edu/abs/2015ARA%26A..53..409W} {53, 409}

\bibitem[\protect\citeauthoryear{{Zhang}, {Blake}  \& {Bergin}}{{Zhang}
  et~al.}{2015}]{ZhangEtal15-cond-front}
{Zhang} K.,  {Blake} G.~A.,   {Bergin} E.~A.,  2015, \mn@doi [\apjl]
  {10.1088/2041-8205/806/1/L7}, \href
  {http://adsabs.harvard.edu/abs/2015ApJ...806L...7Z} {806, L7}

\bibitem[\protect\citeauthoryear{{Zhang} et~al.,}{{Zhang}
  et~al.}{2018}]{Dsharp7}
{Zhang} S.,  et~al., 2018, \mn@doi [\apjl] {10.3847/2041-8213/aaf744}, \href
  {http://adsabs.harvard.edu/abs/2018ApJ...869L..47Z} {869, L47}

\makeatother
\end{thebibliography}

\appendix


\section{Gaseous disc evolution and planet migration}\label{sec:app_numerics}

Here we provide some finer technical detail of planet-disc angular momentum exchange implementation in our code. The total torque $\Lambda_t$ from the planet onto the disc is the integral over the entire disc,
\begin{equation}
    \Lambda_t = \int_{R_{\rm in}}^{R_{\rm out}} \lambda_t(R) \Sigma(R) 2\pi R dR\;.
    \label{lambda_t_int}
\end{equation}
Strictly speaking, the widely used type II torque implementation given by eq. \ref{eq:torque} is well justified for massive planets that open deep gaps in the discs (Type II migration). For less massive planets, type I migration occurs, for which the corotation torques are important \citep[e.g.,][]{PaardekooperEtal10a}. Eq. \ref{eq:torque} then usually under-estimates the migration rate of the planet. A frequent choice for $\Lambda_t$ in this regime are the expressions  from \cite{Paardekooper11-typeI}. In planet formation population synthesis one then also neglects the reverse (planetary) torque on the disc, setting $\lambda_t =0$, arguing that in type I the planet is not very massive and hence can be thought as a test particle. Such an approach is used, for example, by \cite{ColemanNelson14,ColemanNelson16,Bern20-1,Bern20-2}. 

Unfortunately, the abrupt transition from using eq. \ref{eq:torque} for $\lambda_t$ in type II regime to setting it to zero in type I is not appropriate for us here since we aim to model both planet migration and tidal torque effects on the gas {\em and dust} components of the disc. The latter is more strongly influenced by the planet for Stokes numbers not too small \citep[e.g.,][]{DipierroLaibe17}.

Therefore, instead of neglecting the planet torques on the disc in type I regime, we blend the two approaches used in the literature previously. In the type II regime, that is, when $C_{\rm p}\leq 1$, we use eq. \ref{eq:torque}. In Type I regime, when $C_{\rm p}> 1$, instead of setting $\lambda_t$ to zero we write 
\begin{eqnarray}
\label{eq:torque-2}
\lambda_{t}= g \lambda_{\rm out} & \text{ for }
R>a \\
\nonumber \lambda_{t}= g^{-1} \lambda_{\rm in} & \text{ for } R<a\;,
\end{eqnarray}
where $\lambda_{\rm out} = \lambda_t(R > a)$, $\lambda_{\rm in} = \lambda_t(R < a)$ as given in eq. \ref{eq:torque}, and $g$ is a factor of order unity introduced to re-normalise the tidal torque on both sides of the planet to obtain the expected planet migration rate. We find $g$ by computing the integral tidal torque, $\Lambda_t$ from the planet on the disc, and demanding that it equals to that given by the Type I torque $\Lambda_1$ \cite{Paardekooper11-typeI}, with the minus sign, 
\begin{equation}
    g \Lambda_{\rm out} + g^{-1} \Lambda_{\rm in} = -\Lambda_1\;,
    \label{g-factor}
\end{equation}
where $\Lambda_{\rm in}$ and $\Lambda_{\rm out}$ are given by the integral in eq. \ref{lambda_t_int}, except only inside and outside $a$, respectively.
 In practice $g$ is usually between $1$ and $2$. Note that one may think that a symmetric re-normalisation of eq. \ref{eq:torque} is a more logical idea, however we found that in general such an approach requires $g$ that can be either much larger or much smaller than unity. Furthermore, $g$ may even have negative values if type I migration direction is opposite to that predicted to type II. 

Comparing our 1D planet migration treatment with 2D simulations with the public code FARGO3D  \citep{BL_Masset_16_FARGO3D} for a particular disc setup, and a range of disc masses, we found a reasonable agreement. The 1D planet migration time was a factor $f_{1to2}= $1-3 times shorter than the 2D result. The variation in $f_{1to2}$ is mainly due to fact that the 1D treatment cannot capture the type III runaway migration that may occur on the boundary between type I and type II regimes \citep[e.g.,][]{CridaMorbidelli07,CridaBitsch-17-runaway-migration}.

\section{A colder disc in HD163296}\label{sec:App_cold_disc}

As remarked in \S \ref{sec:HD_SS}, we chose a rather hot disc to study in HD163296 by setting $T_0 = 65$~K at $R_0 =100$~AU following \cite{RabEtal20-Temperature-in-gaps}. This temperature is however found at the CO molecule emitting surface, and is larger than $T\approx 25$~K temperature derived for the midplane of the disc in this source by \cite{DullemondEtal20-HD163296}. Here we show that choosing this lower temperature results in even faster planet growth and hence makes our main conclusions even stronger. 

\begin{figure*}
\includegraphics[width=0.99\textwidth]{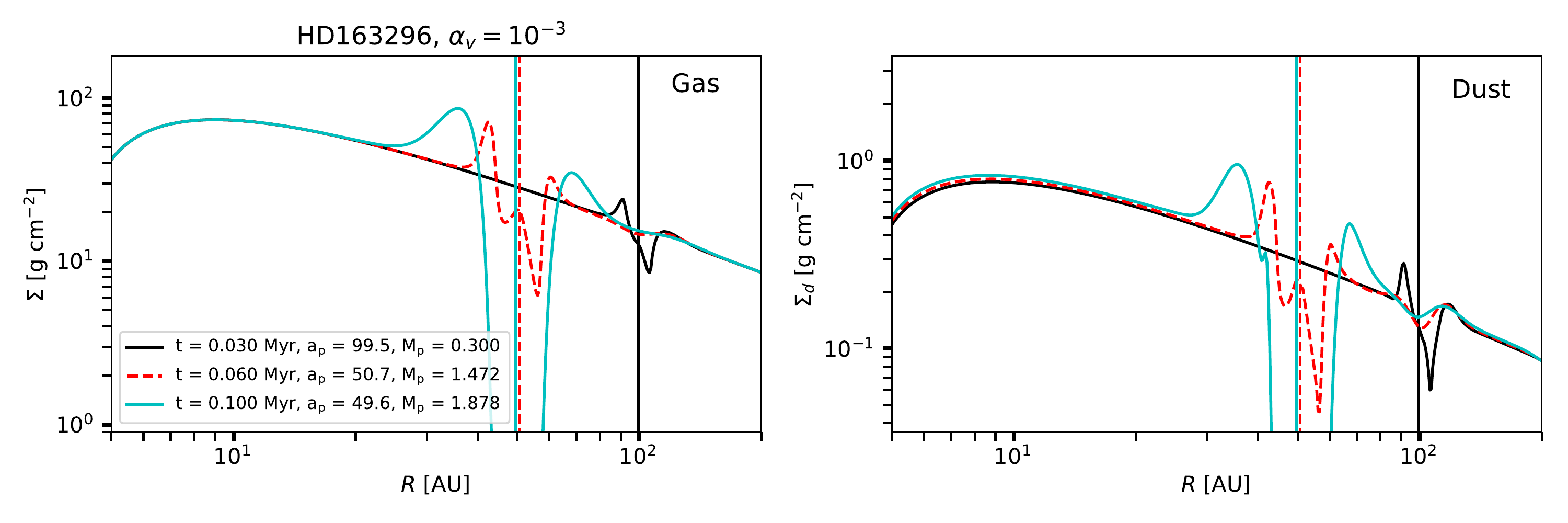}
\includegraphics[width=0.99\textwidth]{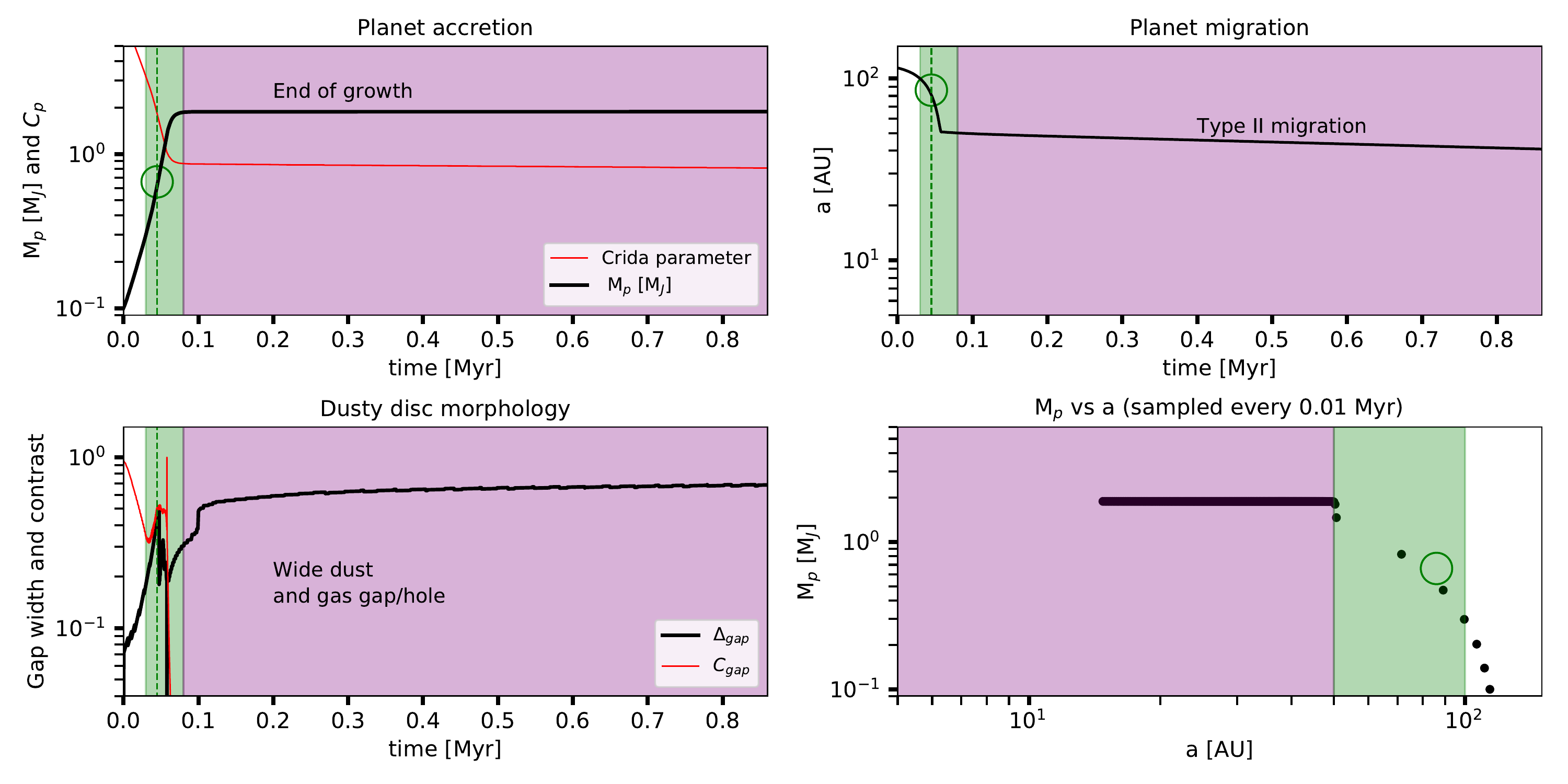}
\caption{Same as Fig. \ref{fig:HD-one-run} but for a colder disc with $T_0 = 25$~K. Note that the planet accretion is faster in this colder disc, therefore it reaches its final mass faster. For the same reason it opens a deep gap and switches to type II migration more rapidly than in the hotter disc studied in Fig. \ref{fig:HD-one-run}.}
\label{fig:T25_hist}
\end{figure*}

Fig. \ref{fig:T25_hist} shows calculation of the disc-planet system evolution repeated in exactly the same way as shown in Fig. \ref{fig:HD-one-run}, but now for a colder disc, with $T_0 = 25$~K. Since gas accretion and gas capture radii are larger in colder discs (cf. eq. \ref{Rgc0}), gas accretion rate is higher in a colder disc too. Therefore, planet runaway to their final masses more rapidly, and hence the time that the gap properties are commensurate with a moderately massive planet becomes yet shorter, that is, less than $0.1$~Myr for the $\alpha_{\rm v} = 10^{-3}$ case presented in Fig. \ref{fig:T25_hist}. For the same reason the planet opens a deep gap in the gas and switches into the type II migration regime earlier and remains at large separations for longer. While retention of the planet at large separation for a few Myr is a welcome result, the gas and dust disc morphology are very different from those observed in HD163296 when the planet is in type II migration regime. Therefore we conclude that our main results are robust with respect to changes in the assumed disc temperature.

\bsp	
\label{lastpage}
\end{document}